\documentclass[floatfix,aps,prl,twocolumn,showpacs,superscriptaddress,10pt,longbibliography]{revtex4-2}

\usepackage{amsmath,amsfonts,amssymb}
\usepackage{graphicx}
\usepackage{epstopdf}
\usepackage{bm}
\usepackage{hyperref}
\usepackage{orcidlink}

\usepackage[utf8]{inputenc}
\usepackage[T1]{fontenc}
\usepackage[english]{babel}

\IfFileExists{newtxtext.sty}
 {\usepackage{newtxtext,newtxmath}}
 {\IfFileExists{stix.sty}
  {\usepackage{stix}}
  {\IfFileExists{mathptmx.sty}
  {\usepackage{mathptmx}}{} } }

\usepackage{xcolor}

\begin{document}








\title{Reproducible nucleation and control of stable quantum vortex rings in Bose--Einstein condensates
}

\author{Giorgia Iori\,\orcidlink{0009-0005-2396-6689}}
\affiliation{Dipartimento di Fisica ``Aldo Pontremoli'', Universit\`a degli Studi di Milano, via Celoria 16, I-20133 Milano, Italy}
\author{Klejdja Xhani\,\orcidlink{0000-0003-0713-8523}}
\affiliation{Dipartimento Scienza Applicata e Tecnologia, DISAT, Politecnico di Torino,  I-10129, Torino, Italy}
\author{Woo Jin Kwon\,\orcidlink{0000-0001-9773-4024}}
\affiliation{Dipartimento di Fisica ``Aldo Pontremoli'', Universit\`a degli Studi di Milano, via Celoria 16, I-20133 Milano, Italy}
\affiliation{Department of Physics, Ulsan National Institute of Science and Technology (UNIST), 44919 Ulsan, Republic of Korea}
\author{Davide Emilio Galli\,\orcidlink{0000-0002-1312-1181}}
\email{davide.galli@unimi.it}
\affiliation{Dipartimento di Fisica ``Aldo Pontremoli'', Universit\`a degli Studi di Milano, via Celoria 16, I-20133 Milano, Italy}
\author{Luca Galantucci\,\orcidlink{0000-0002-3435-4259}}
\email{luca.galantucci@cnr.it}
\affiliation{Istituto per le Applicazioni del Calcolo, Consiglio Nazionale delle Ricerche, via dei Taurini 19, I-00185 Roma, Italy}
\affiliation{School of Mathematics, Statistics and Physics, Newcastle University, NE1 7RU Newcastle upon Tyne, United Kingdom}

\begin{abstract}
We propose and numerically validate an experimentally feasible on--demand protocol for the nucleation and manipulation of stable quantum vortex rings in trapped Bose-–Einstein condensates.
The method relies on sweeping a laser--sheet barrier that locally constricts the superflow and triggers vortex--ring formation. By tuning the barrier height and width, and by scanning the barrier velocity, we identify the onset of periodic generation of vortex rings above the critical velocity and achieve direct, deterministic control over the ring nucleation position, radius, and hence propagation speed. After its formation, ad--hoc optical potentials are applied to reshape the vortex ring, creating clean Kelvin-wave excitations. Our results provide a foundation for systematic studies of three-dimensional vortices in atomic superfluids and open the door to tailored vortex dynamics and interactions, enabling controlled access to quantum turbulence.
\end{abstract}

\maketitle

\textit{Introduction}.
Coherent filamentary structures are ubiquitous across physical systems, from classical fluids and plasmas to optical fields and quantum matter. Their dynamics play a key role in determining the macroscopic properties of nematic liquid crystals \cite{chuang-etal-1991}, plasmas (from astrophysical flows \cite{priest-forbes-2007,che-etal-2011,cirtain-etal-2013} to nuclear fusion devices) and optical beams \cite{dennis-etal-2010,berry-dennis-2012}, as well as classical \cite{farge-etal-2001,marusic-monty-2019} and quantum fluids \cite{donnelly-1991,barenghi-donnelly-vinen-2001,Barenghi_Skrbek_Sreenivasan_2023}.

In fluid systems, these coherent structures correspond to vortices: localized regions of concentrated vorticity that organize the surrounding flow and play a central role in turbulence. Through nonlinear interactions and vortex stretching, vortices drive the turbulent cascade \cite{richardson-1922}, redistributing kinetic energy across length scales according to the Kolmogorov spectrum \cite{kolmogorov-1941}, observed in both classical and quantum turbulent flows \cite{maurer-tabeling-1998,barenghi-lvov-roche-2014,baggaley-barenghi-shukurov-sergeev-2012,nore-abid-brachet-1997}.
At the smallest scales, vortices are ultimately responsible for energy dissipation, either through viscosity in classical fluids or via mutual friction \cite{galantucci-etal-2023} and Kelvin-wave cascades \cite{vinen-2001-decay,kivotides-etal-2001,lvov-nazarenko-2010,kozik-svistunov-2004,galantucci-etal-2023} in quantum fluids.

The nature of vortices is fundamentally dictated by whether the fluid is classical or quantum. In classical fluids, vortices are unconstrained in strength, size, and geometry and are dissipated by viscosity. In contrast, in quantum fluids —such as superfluid helium \cite{Barenghi_Skrbek_Sreenivasan_2023}, ultracold atomic gases \cite{white-etal-2014}, polariton condensates \cite{panico-etal-2023}, and the interior of neutron stars \cite{haskell-melatos-2015} — vorticity is confined to discrete, effectively one-dimensional vortex lines with quantized circulation and fixed core size \cite{onsager-1949,feynman-1955,vinen-1961}.
These vortex lines are topological defects of the order parameter and constitute stable excitations in homogeneous quantum fluids at zero temperature. Their discrete filamentary nature, together with the exceptional experimental tunability of Bose--Einstein condensates, makes ultracold gases an ideal platform for investigating fundamental vortex dynamics, including vortex reconnections \cite{serafini-etal-2017,galantucci-baggaley-parker-barenghi-2019,villois-etal-2020}, vortex–boundary interactions \cite{stagg-parker-barenghi-2017}, and Kelvin-wave excitations \cite{krstulovic2023kelvin}.
The relevance of these studies goes beyond quantum fluids as,
due to striking observed similarities between quantum and classical turbulence
\cite{maurer-tabeling-1998,stalp-skrbek-donnelly-1999,lamantia-skrbek-2014,muller-etal-2021}, 
the suggestive idea that turbulent flows, both classical and quantum, may be described in terms of the 
mutually-interacting dynamics of discrete and thin vortex filaments of fixed circulation has been put forward
\cite{polanco-etal-2021,galantucci-2025}.

Despite this progress, a central experimental challenge is the on-demand, reproducible generation of individual three-dimensional (3D) quantum vortices, with control over their geometry, dynamics, and lifetime. While in two dimensions
the topic has been successfully tackled in both Bose and Fermi atomic superfluids employing the so-called chopstick method 
\cite{samson-etal-2016,kwon-etal-2021},
in three dimensions, this remains challenging because additional degrees of freedom (e.g., Kelvin-wave excitations) can reduce vortex lifetime and affect stability. 
In 3D condensates, vortices are currently generated either stochastically via the Kibble–Zurek mechanism \cite{serafini-etal-2015}, through Josephson junction dynamics \cite{burchianti-etal-2018}, or by driving the condensate with quantum pistons \cite{mossman-etal-2018}. In all these approaches, the absence of precise control over vortex number, geometry, and reproducibility hinders systematic studies of vortex interactions. Numerical studies have provided relevant insight into 3D vortex nucleation in trapped condensates \cite{piazza-etal-2011,xhani-etal-2020,Xhani_njp2020,singh-etal-2025} in the context of the critical velocity determination, vortex-ring dynamics and classification of generated solitary waves (vortex rings or rarefaction pulses \cite{jones-roberts-1982}), related to small and unstable vortex rings (radii smaller than 5 times the coherence length). Nevertheless, many aspects of the direct control of stable vortex rings remain elusive.

In this work, we numerically demonstrate a protocol for the controlled nucleation and manipulation of stable and reproducible quantum vortex rings in trapped Bose--Einstein condensates.
By employing a moving laser sheet, we achieve precise control over the nucleation position, size, shape, and propagation velocity of vortex rings, as well as their number and generation frequency. Breaking the axial density symmetry by applying localized beams in the downstream, we further enable controlled excitation of Kelvin waves, paving the way for systematic studies of quantum vortex excitations.
 
\begin{figure*}[t]
    \centering
    \includegraphics[width=\textwidth]{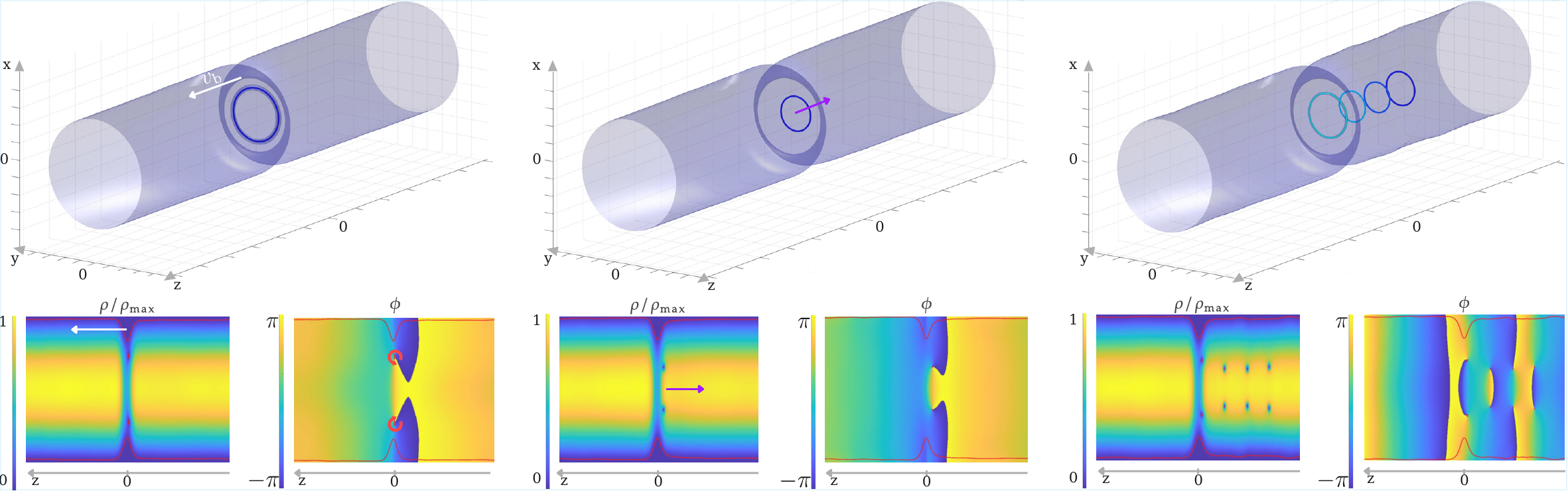}
    \caption{Stages of vortex ring nucleation and evolution, shown via BEC density isosurfaces at 5\% of the maximum density and density $\rho$ -phase $\phi$ projections on the plane $y=0$ at different times in the reference frame where the barrier is stationary.
    Left panel: Ring generation triggered by the moving barrier (white arrow); red arrows in the phase snapshot indicate the circulation of the velocity field. 
    Central panel: Shrinking across the barrier: the ring propagates opposite to the barrier motion (purple arrow) while its radius decreases. 
    Right panel: Propagation in the bulk and periodic generation: the first ring reaches an asymptotic regime with constant radius and velocity; meanwhile, additional rings are generated periodically and eventually reach the same asymptotic properties. In the isosurfaces, the vortex rings are represented by circles displayed in different shades of blue. 
    The red curves on the 
    density-phase plots show the low-density isocontour at $2.5\%$ of the maximum density, marking the condensate edge and the density depletions corresponding to vortex cores. 
    }
    \label{fig:fig1}
\end{figure*}

\textit{Problem and approach}.
We model the dynamics of trapped Bose--Einstein Condensates employing the mean-field Gross--Pitaevskii framework \cite{pitaevskii-stringari-2003} according to which the
order parameter $\Psi(\mathbf{x},t)$ of the system is governed by the following equation 
\begin{equation}
\displaystyle
i \hbar\frac{\partial \Psi}{\partial t} = -\frac{\hbar^2}{2 m}\nabla^2\Psi + g |\Psi|^2\Psi + V \Psi -\mu \Psi \; \; , 
\label{eq:GP}
\end{equation}
where $m$ is the mass of boson considered, $g=4 \pi \hbar^2 a_s / m$ is the two-body interaction strength, $a_s$ being the $s$-scattering length, 
$V=V(\mathbf{x},t)$ is the externally applied potential and $\mu$ is the chemical potential, related to the number of particles $N$ and $a_s$ via 
the normalisation condition $\displaystyle \int \!|\Psi|^2 d\textbf{x} = N$. The potential $V=V_h + V_b$ is composed of the static cylindrical harmonic potential
$\displaystyle V_h(\textbf{x}) = \frac{1}{2} m \omega_\perp (x^2 + y^2)$, where $\omega_\perp = 2 \pi 150 \rm Hz$ is the radial trapping frequency, 
and the time--dependent planar barrier potential 
$\displaystyle V_b(\textbf{x},t)=V_b(z,t)=\displaystyle V_0 e^{-\frac{(z-z_b(t))^2}{2\sigma_b^2}}$, 
where $V_0$ is the tunable barrier intensity, $\sigma_b$ the width of the barrier 
and $z_b(t)$ is the controllable position of the center of the barrier. 
The potential $V_b$ describes the impact on the condensate of the axial sweep of a laser sheet of width $\sigma_b$. In our study the barrier moves with a gradually ramped-up in time 
velocity $v_b (t)=v_b^\mathrm{max}\tanh (t/\tau_b)$ \cite{SuppMat} from an initial axial position $z_0$. 
We choose a cylindrical condensate, with no axial trapping ($\omega_z=0$), in order to study in an idealized
setting the dynamics occurring in the central region of an experimentally realizable cigar-shaped condensate. The radial size of the condensate far from the barrier is 
$R_{\perp_\infty}=\sqrt{2\mu/(m\omega_\perp^2)}$ in the limit of large $N$ which we satisfy as we choose $N\sim 1.5 \times 10^5$ pairs of $^6$Li atoms, implying 
thus $m=2\times 10^{-26} \rm kg$ and $a_s=5.3 \times 10^{-8} \rm m$. Details on the numerical simulations 
are reported in the Supplemental Material \cite{SuppMat} and hereafter all quantities will be indicated in dimensionless units, the units of length, time and energy
being respectively $\ell_\perp=\sqrt{\hbar/(m \omega_\perp)}$, $\tau_\perp = \omega_\perp^{-1}$ and $\epsilon_\perp=\hbar \omega_\perp$ \cite{Iori-thesis-2025}.

The vortex nucleation process is as follows.
At time $t=0$, we start moving the barrier towards the positive $z$-direction from the initial position $z_0=-R_{\perp_\infty}$, which we choose arbitrarily without loss of generality. 
The motion of the barrier is equivalent (with a Galileian transformation) to forcing the fluid (the condensate) 
to flow in a constricted geometry, accelerating its speed. 
In fact, the presence of the moving barrier narrows the radial size of the condensate, which at $z=z_b(t)$ is equal to 
$R_{\perp_b}=\sqrt{2(\mu-V_0)} < R_{\perp_\infty} = \sqrt{2\mu}$
(in non-dimensional units),  
and depletes the particle density $n$ of the condensate 
$n_b(r)=n(r,z)\vert_{z=z_b(t)}=(\mu - V_0 - V_h(r))/C < (\mu - V_h(r))/C = n_\infty(r)$, where $r$ is the radial coordinate, $C=4 \pi N a_s/\ell_\perp$ is the
non dimensional interaction strength \cite{SuppMat} and $n_\infty(r)$ is the radial density profile far from the barrier. 
The conservation of the mass flux thus implies that at the barrier, the average axial relative velocity $\bar{v}_b^z$ between the condensate and the barrier is larger than the average relative velocity $\bar{v}^z_\infty$ far from the barrier itself, and as we move radially towards the boundaries of the condensate the axial velocity increases, due to the presence of quantum pressure effects \cite{piazza-etal-2011}. 
%
Therefore, the maximum relative velocity between the barrier and the condensate occurs at $z=z_b(t)$ near the boundary ($r\sim R_{\perp_b}$). 
Density depletion there lowers the local sound speed $c$, closely related to the critical velocity for vortex nucleation either via the Landau criterion ($v_c=c$) \cite{berloff-roberts-2000,stagg-etal-2014,kwon2015critical} or via the triggering of a transonic transition determined by the change of the nature of the differential equation governing the flow (elliptic to hyperbolic \cite{frisch-etal-1992,musser2019starting,frisch2024superflow}). Exceeding this critical velocity causes the nucleation of a vortex ring at the barrier edge, as shown in Fig.~\ref{fig:fig1} (left).

\begin{figure}[htbp]
    \centering
    \includegraphics[width=1.0\columnwidth]{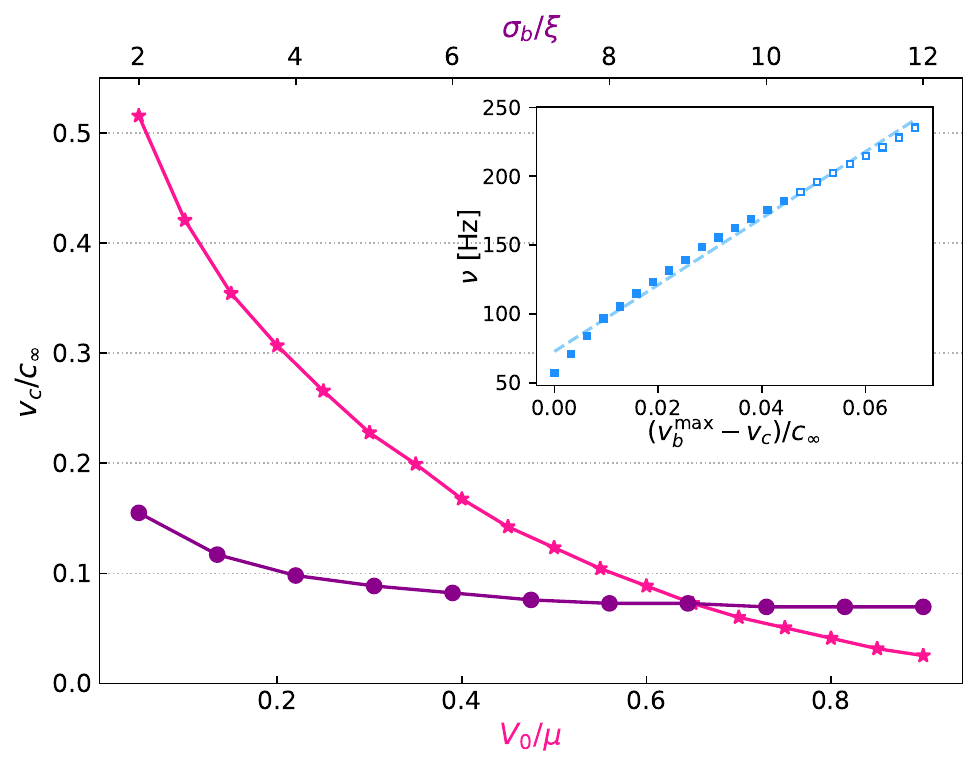}
    \caption{Critical velocity $v_c$ for vortex ring nucleation as a function of barrier height $V_0/\mu$ (pink stars, fixed $\sigma_b=5\xi$) and of barrier width $\sigma_b$ (purple circles, fixed $V_0=0.6\mu$).
    Inset: vortex nucleation frequency $\nu$ at fixed $V_0=0.6\mu$ and $\sigma_b=5\xi$ as a function of barrier velocity $v_b^\mathrm{max}>v_c$ (blue squares), spanning from $v_c$ to $1.78v_c$; the dashed line represents a linear fit of the data. Empty squares represent the cases where we observe leapfrogging events. Velocities are expressed in units of the density-averaged speed of sound $c_\infty=\sqrt{\mu/2}$.}
    \label{fig:vcrit}
\end{figure}

As the tunable barrier intensity $V_0$ is increased, the constriction of the flow is more pronounced, leading to a narrower flow cross-section and thus to a stronger acceleration of the flow at the barrier. Hence, a smaller velocity of the barrier is needed in order to exceed the local critical velocity and nucleate the first vortex ring. We indicate the smallest velocity of the barrier $v_b^\mathrm{max}$ at which we observe vortex nucleation as the critical velocity $v_c$ and report its behaviour as a function of $V_0/\mu$ in the main panel of Fig.~\ref{fig:vcrit}, showing indeed a decrease of $v_c$ for increasing $V_0$. A similar behaviour of $v_c$ is observed 
experimentally in two-dimensions \cite{kwon2015critical}. We also study the variation of $v_c$ for increasing barrier width $\sigma_b$ (Fig.~\ref{fig:vcrit}, main panel; top axis), observing 
that the critical velocity reaches a plateau once the width has reached the size of a vortex core $a_0 \sim 5\xi$, $\xi=1/\sqrt{2\mu}$ being the healing length on the trap axis, 
as observed in homogeneous condensates \cite{huepe-brachet-2000,stagg-etal-2014}. 


Once nucleated, the vortex undergoes a dynamics whose main stages are reported in
Fig.~\ref{fig:fig1}, with three-dimensional renderings (top row) and two-dimensional density and phase plots
on the horizontal $y=0$ plane (bottom row); vortices are tracked with an algorithm based on the pseudo-vorticity 
\cite{rorai-etal-2016,villois-etal-2016} (top) and identified by density depletions and phase defects (bottom). 
We observe indeed the nucleation of a vortex ring near the boundary in the barrier region (Fig.~\ref{fig:fig1}, left panel). Its nucleation corresponds to an injection of energy (mainly kinetic) into the condensate resulting from the work performed by the barrier on the condensate itself (see End Matter). The nucleation process, in fact, strongly breaks the upstream-downstream density symmetry at the barrier, generating a force performing the work and injecting the energy into the condensate. The dynamics of the vortex ring instead is driven by its self-induced velocity, while its shrinking arises from density gradient forces or, equivalently, from energy conservation constraints (Fig.~\ref{fig:fig1}, central panel). Far from the barrier, the ring propagates at a constant velocity, preserving its shape (Fig.~\ref{fig:fig1}, right panel).

It is worth noting that the nucleated vortex ring travels  opposite to the barrier motion, unlike the vortices generated by the flow around a single obstacle (\textit{e.g.} a disc in two-dimensions) \cite{sasaki-etal-2010,kwon2016observation}. Instead, its propagation direction coincides with that of vortices generated between multiple obstacles \cite{aioi2011controlled}.

Once $v_b \gtrsim v_c$, we observe periodic generation of vortex rings, each undergoing the same dynamical evolution [Fig.~\ref{fig:fig1} (right)], the frequency of nucleation increasing approximately linearly with the barrier velocity, as reported in the inset of Fig.~\ref{fig:vcrit}. When the frequency is sufficiently large (empty symbols in Fig.~\ref{fig:vcrit} (inset)), the distance between vortices is small enough to trigger a leapfrogging dynamics \cite{Galantucci_Sciacca_Parker_Baggaley_Barenghi_2021,singh-etal-2025}.


\begin{figure}[htbp]
    \centering
    \includegraphics[width=1.0\columnwidth]{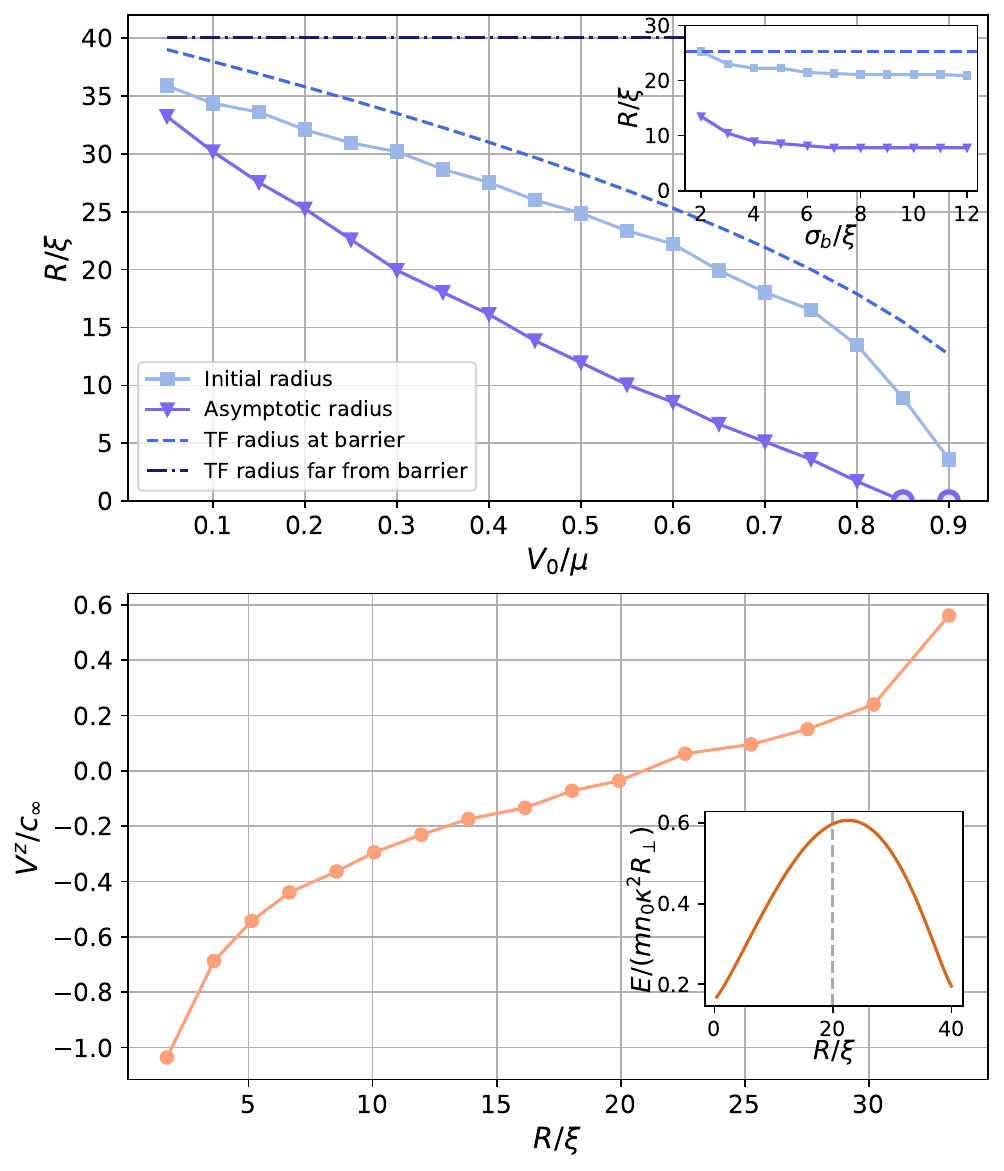}
    \caption{Top panel: Initial (light--blue squares) and asymptotic (lilac triangles) vortex ring radii $R$ at critical barrier velocity as a function of the barrier intensity $V_0/\mu$ with fixed $\sigma_b=5\xi$. Inset: $R$ as a function of barrier width $\sigma_b$ with fixed $V_0=0.6\mu$. 
    Open lilac circles indicate asymptotic radii set to zero as the rings self-annihilate before reaching an asymptotic regime. 
    Bottom panel: Asymptotic vortex ring axial velocity $V^z$ as a function of its asymptotic radius $R$ reported in the top panel. Inset: Kinetic energy $E$ of a cylindrical 
    condensate in the presence of a vortex ring as a function of the ring's radius $R$; the dashed gray line at $R=0.5R_{\perp_\infty}$ marks where the slope would change sign in a homogeneous system. Here $n_0$ indicates the particle number density on the cylinder axis.}
    \label{fig:radius}
\end{figure}

In order to study reproducibly vortex interactions, it is essential to be able to predict the size of the generated vortex rings. 
In Fig.~\ref{fig:radius} (top), we report the initial (light-blue) and asymptotic (lilac) radii $R$ of the generated vortex rings as a function of $V_0/\mu$ and their behaviour as we vary the barrier width $\sigma_b$ for $V_0/\mu=0.6$ (inset). 
The initial radius clearly follows the radial width of the condensate at the barrier $R_{\perp_b}$
indicated with a dashed curve in Fig.~\ref{fig:radius} (top). The difference between initial and asymptotic values increases with $V_0/\mu$: 
indeed larger $V_0$ implies larger density gradients, which are responsible for the shrinking of the vortex ring. The inset of Fig.~\ref{fig:radius} (top)
shows again that the width of the barrier $\sigma_b$ only impacts when smaller than $a_0$. 

The other key element to 
investigate vortex interactions deterministically is the vortex ring velocity. While in a homogeneous and 
unbounded condensate the expression relating radius $R$ and velocity $V^z$ is straightforward, namely $V^z \sim 1/R$ for $R$ sufficiently larger than $\xi$ 
\cite{roberts-grant-1971}, the existence of boundaries \cite{mason-berloff-fetter-2006} 
and/or the presence of inhomogeneous density backgrounds \cite{mason-berloff-2008,komineas-papanicolaou-2002} introduce additional physical effects
whose nonlinear combination governs the speed and the direction of the generated vortex ring. The density depletion near the boundary of a condensate 
indeed shifts the position of the image vortex \cite{mason-berloff-fetter-2006} which together with the density gradient dictates the 
velocity of the vortex ring. The measured asymptotic velocity of the nucleated rings is reported in Fig.~\ref{fig:radius} (bottom) as a function of 
the ring radius $R$. Three non-trivial characteristics emerge clearly: first, there is a direction inversion of the ring's motion; second, 
this change of velocity sign does not occur at $R/\xi=20$ ($R/R_{\perp_\infty}=0.5$); third, the velocity magnitude is not symmetrical w.r.t. $V^z=0$. 
The change of direction of the vortex ring motion is observed in both classical and quantum confined fluids
\cite{Galantucci_Sciacca_Parker_Baggaley_Barenghi_2021,mason-berloff-2008,komineas-papanicolaou-2002}, while the asymmetric characteristic 
emerging in Fig.~\ref{fig:radius} (bottom) is peculiar to BECs and arises from their compressibility and from the density modulation which
can be achieved employing external potentials (for instance, in a classical homogeneous two-dimensional channel, the velocity
is reversed exactly at $R/R_{\perp}=0.5$ \cite{SuppMat}). The velocity dependence on $R$ and in particular the value of
$R$ for which vortex $V^z=0$ can be determined, recalling that, as in classical fluid dynamics, quantum vortex rings and
pairs follow Hamiltonian mechanics both in homogeneous \cite{jones-roberts-1982} and inhomogeneous condensates \cite{mason-berloff-2008,komineas-papanicolaou-2002}.
Thus the velocity of vortex rings is given by the following group-velocity relation 

\begin{equation}
\displaystyle
V^z=\frac{\partial E}{\partial p_z} = \frac{\partial E/\partial R}{\partial p_z/\partial R} \,\, ,
\label{eq:U_hamilton}
\end{equation}

\noindent
where $\displaystyle E=\int\frac{1}{2}\rho v^2 d\mathbf{x}$ is the kinetic energy induced by the vortex ring, 
$\displaystyle p_z=\int\rho v_z d\mathbf{x}$ is the momentum of the fluid in the axial direction, $v$ and $v_z$
are the fluid velocity magnitude and its axial component, respectively.
We compute analytically $E$ and $p_z$ in a cylindrical harmonic trap without the barrier and in the presence of a vortex ring, in order 
to study the asymptotic ring motion. In cylindrical coordinates, with no dependence on the azimuthal angle due to the symmetry of the problem, we approximate the condensate density $\rho(\mathbf{r})$ at a generic point $\mathbf{r}=(r,z)$
interpolating asymptotic forms of the density close and far away from the vortex located at $\mathbf{R}=(R,z_R)$ \cite{jackson-etal-1999}, $R$ being the vortex ring radius and $z_R$ its position along the cylinder axis. This yields 
$\displaystyle \rho \approx m n_0 \left [ \frac{|\mathbf{r}-\mathbf{R}|^2}{|\mathbf{r}-\mathbf{R}|^2 + \xi^2}\right ] 
\left [ 1-\left ( \frac{r_\perp}{R_\perp} \right )^2 \right ]$, $n_0$ being the particle number density on the cylinder axis and $r_\perp$
distance from the latter. We employ a local velocity approximation according to which the velocity $\mathbf{v}$ of the fluid is
only dictated by the closest vortex line element, namely $\mathbf{v}=\bm{\kappa} \times (\mathbf{r}-\mathbf{R}) /(2\pi|\mathbf{r}-\mathbf{R}|^2)$,
where $\bm{\kappa}$ is the oriented vortex line element parallel to the vorticity with magnitude equal to the quantum of circulation
$\kappa=h/m$ \cite{SuppMat}. The behaviour of $E$ as a function of $R$ is reported in Fig.~\ref{fig:radius} (bottom, inset) showing a
slope sign switch for the same $R$ where $V^z$ changes its sign. Together with a monotonous behaviour of $p_z(R)$ having a negative 
derivative for all $R$ (see End Matter), it explains the behaviour of the vortex ring velocity. It is worth noting that the approximations
employed in the analytical calculations are not able to describe reliably $E$ and $p_z$ for small radii. 

\begin{figure}[htbp]
    \centering
    \includegraphics[width=1.0\columnwidth]{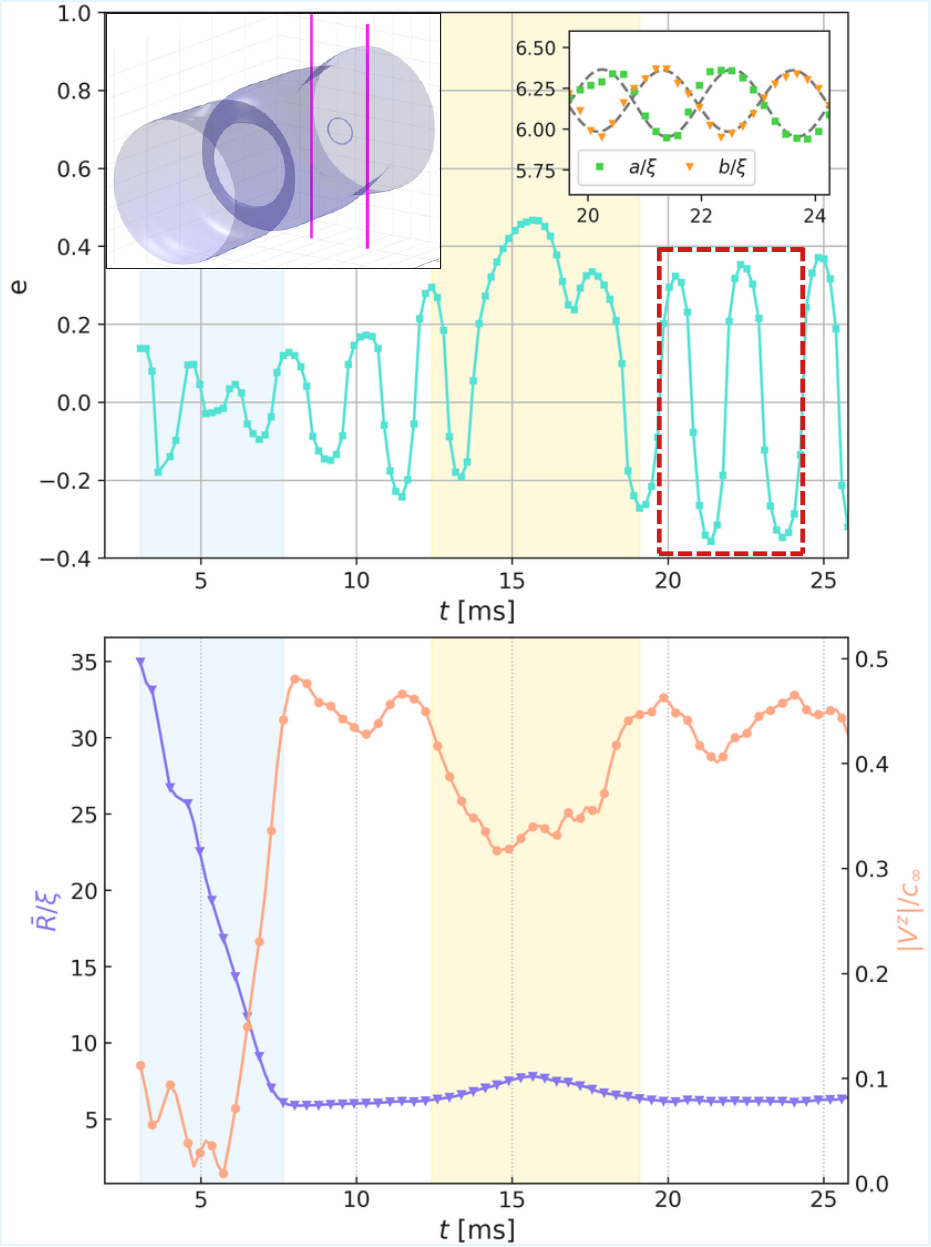}
    \caption{Triggering of $m=2$ Kelvin waves on a vortex ring. Top panel: Time evolution of the eccentricity of the vortex ring projection on the $x$-$y$ plane. Left inset: BEC density iso-surface, with the two local vertical potentials $V_i(\mathbf{x})$ represented as pink columns. Right inset: when the ring approaches the obstacles, they break the symmetry and excite oscillations of its semiaxes $a$ and $b$; the dashed gray curve is a sinusoidal fit that matches the theoretical frequency of the $m=2$ Kelvin wave. Bottom panel: Time evolution of the mean radius $\bar{R}=(a+b)/2$ (left axis, lilac triangles) and the velocity $V^z$ of the vortex ring (right axis, light-pink circles). The blue area represents the region where the vortex interacts with the moving barrier, while the yellow area the region where it is affected by the vertical obstacles. 
    }
    \label{fig:fig4}
\end{figure}
To expand the variety of vortex interaction 
and generate excitations along the vortex rings,
we include additional features in our platform. In particular, downstream with respect to the barrier, we shine two laser beams parallel to the $y$ direction 
[Fig.~\ref{fig:fig4} (top, left inset)] which 
correspond to local vertical potentials 
$\displaystyle V_i(\mathbf{x})=V_c \exp{\{-\left[(x-x_i)^2+(z-z_c)^2\right]/2\sigma_c^2\}}$, where $\sigma_c=10\xi$, $V_c/\mu=3/8$, $z_c=-10$,
$x_i=\pm 3$. The presence of these obstacles breaks the axial symmetry of the system 
and triggers the generation and propagation of almost planar Kelvin waves along the vortex rings [Fig.~\ref{fig:fig4} (top)]. 
When the nucleated vortex rings approach the local potentials $V_1$ and $V_2$, 
between $t=10$ and $t=20$, $|z_R-z_c|< 3 \sigma_c$ ($z_R$ being the axial coordinate of the vortex ring), 
the size of BEC and its density are decreased by the potentials, resulting in an increase of the average vortex radius, as shown in [Fig.~\ref{fig:fig4} (bottom panel)]; this increase of the radius of the ring slows its motion, as reported in [Fig.~\ref{fig:fig4} (bottom panel)], consistent with Fig.~\ref{fig:radius} (bottom). 
The expansion of the ring is not axially symmetric, the vortex increases its size in the $x$ direction where the BEC is reduced by $V_i$, as shown from
the evolution of the eccentricity $e$ in [Fig.~\ref{fig:fig4} (top)], where $e=\sqrt{1-(b/a)^2}$ if $a\ge b$ and $e=-\sqrt{1-(a/b)^2}$, $a$ and $b$ being the ellipse semiaxes in the $x$ and $y$ directions respectively. The vortex ring hence becomes elliptical, corresponding to a Kelvin wave of mode $m=2$, which then propagates 
[Fig.~\ref{fig:fig4} (top, right inset)]. The oscillation frequency of the $m=2$ Kelvin wave mode is consistent with previous analytical results \cite{Pocklington-1895}. 
Furthermore, by employing a single off--centered local vertical obstacle $V_i$, we intentionally break the linear propagation of the vortex ring and trigger an instability that drives it to crash onto the condensate boundary, highlighting the versatility of our 3D vortex-manipulation protocol (see End Matter).

In conclusion, we present a protocol for the deterministic and reproducible generation of vortex rings in Bose–Einstein condensates that is experimentally realizable.
By translating a planar optical sheet, our procedure directly controls the radius of the nucleated rings, and hence their velocity, and emission frequency, enabling interactions between successively emitted rings 
(\textit{e.g.} leapfrogging \cite{Galantucci_Sciacca_Parker_Baggaley_Barenghi_2021,singh-etal-2025}). In addition, we show that by employing additional localised
laser beams, we are able to change the shape, the radius and the velocity of the ring, and to generate Kelvin waves on the ring itself. This comprehensive control of the generation and
manipulation of vortex rings in experimentally achievable conditions could be highly relevant for the reproducible 
experimental study of quantum vortex--vortex interactions (including reconnections \cite{serafini-etal-2017}), and the interplay between quantum vortices and boundaries (highly non trivial with respect to the classical incompressible counterpart), and for the direct observation and investigation 
of quantum vortex-line excitations such as Kelvin waves without the use of particle tracers \cite{fonda-etal-2014,peretti-etal-2023,minowa2025direct} trapped on vortices.
The results we present 
may have an impact beyond quantum fluids, regarding fluid dynamics in general. Similarities between classical and
quantum turbulence have suggested that turbulent flows, both classical and quantum, may be depicted on the basis of collective and interactive dynamics of
thin vortex filaments of fixed circulation \cite{polanco-etal-2021,galantucci-2025}: the study of quantum vortex interactions could thus lead to the enhancement
of the understanding of the still open issues in classical turbulence.

\acknowledgments
L.G. acknowledges fruitful discussions with Carlo Barenghi and Giovanni Franzina. D.E.G. and G.I. acknowledge fruitful discussions with Christian Apostoli. We acknowledge computational resources provided by INDACO Platform, which is a project of high performance computing at Università degli Studi di Milano. This work was supported by the European Research Council (ERC) under grant agreement No. 101076129, and the National Research Foundation of Korea (NRF) grant funded by the Korea government (MSIT) (RS-2024-00348882).
\bibliography{mybib5.bib}
\onecolumngrid
\newpage
\begin{center}
 \textbf{\large End Matter}
\end{center}
\twocolumngrid

\appendix

\section{Momentum of vortex ring}

\begin{figure}[htbp]
    \centering
    \includegraphics[width=1.0\columnwidth]{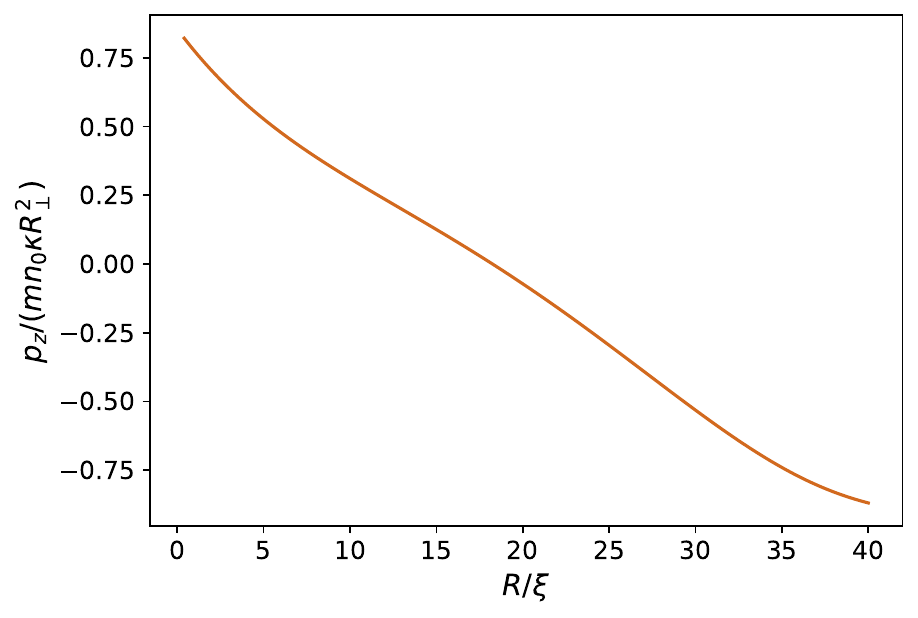}
    \caption{Axial momentum $p_z$ of a cylindrical 
    condensate in the presence of a vortex ring as a function of the ring's radius $R$.}
    \label{fig:momentum}
\end{figure}

The axial momentum of the BEC is defined as 
\begin{equation}
\displaystyle
p_z=\frac{1}{2 i}\int \left ( \Psi^*\frac{\partial \Psi}{\partial z}-\Psi\frac{\partial \Psi^*}{\partial z}\right ) d\mathbf{r} \, \, .
\label{eq:momentum_z_psi}
\end{equation}
If we use the hydrodynamical formulation where $\Psi=\sqrt{n}e^{i\phi}$, where $n=\rho/m$ is the particle density number and $\phi$ the phase such that 
the velocity $\mathbf{v}=\nabla \theta$ then 
\begin{equation}
\displaystyle
p_z=\int \rho(\mathbf{r}) v_z(\mathbf{r}) d\mathbf{r} \, \, ,
\label{eq:momentum_z_hydro}
\end{equation}
where $v_z$ is the axial velocity of the fluid induced by the vortex ring. We compute analytically the integral as a function of the radius
$R$ of the vortex ring \cite{SuppMat} and obtain the result reported in Fig.~\ref{fig:momentum}, where we observe that $p_z$ is a decreasing function
of $R$ in the whole range. Hence, the velocity of the vortex ring reverses its sign when $\partial E/\partial R=0$, \textit{i.e.} for $R\sim 23\xi$.

\section{Energy analysis}
\begin{figure}[htbp]
    \centering
    \includegraphics[width=1.0\columnwidth]{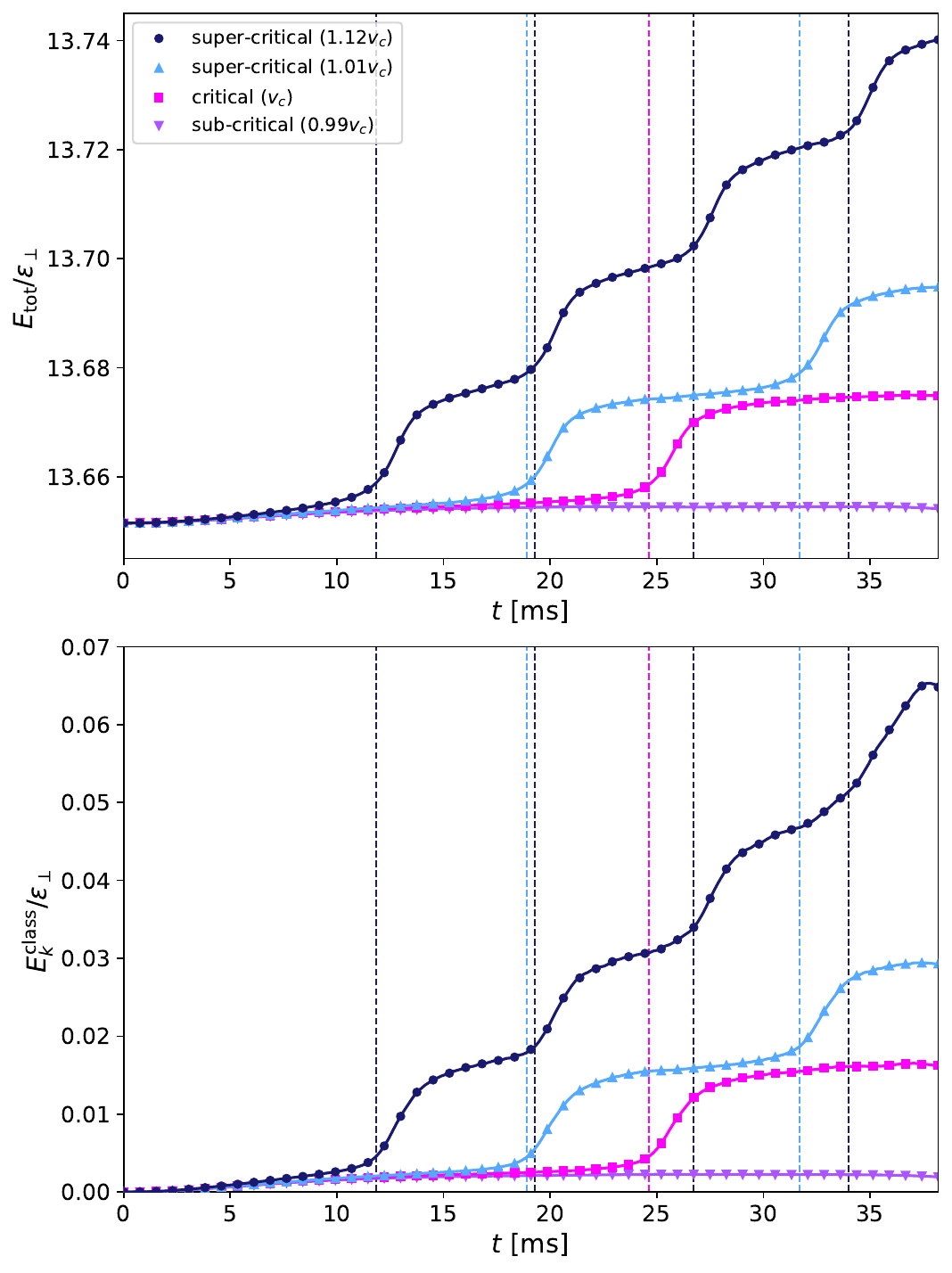}
    \caption{Energy analysis during time evolution, for $V_0/\mu=0.6$ and $\sigma_b=2\xi$. Comparison between the critical case (barrier velocity $v_b^\mathrm{max}=v_c$, pink squares), a sub-critical case ($v_b^\mathrm{max}=0.99v_c$, lilac triangles) and two super-critical cases ($v_b^\mathrm{max}=1.01v_c$, light-blue triangles, and $v_b^\mathrm{max}=1.12v_c$, dark-blue circles). Top panel: Total energy $E_\mathrm{tot}$ of the superfluid, computed as the sum of kinetic, potential and interaction energy contributions. Bottom panel: Temporal evolution of the classical kinetic energy $E_k^\mathrm{class}$.
    Vertical dashed lines mark the time instants when vortex rings are nucleated.}
    \label{fig:energy}
\end{figure}

The total energy of the system is given by the Gross--Pitaevskii energy functional
\begin{equation}
    \label{eq:total_energy}
    E_\mathrm{tot} = \int_\Omega d\Omega \biggl[\frac{1}{2}|\nabla\Psi|^2+V|\Psi|^2+\frac{g}{2}|\Psi|^4 \biggr]
\end{equation}
where $\Omega$ denotes the computational domain corresponding to the numerical grid used to represent the system. 
By analysing the time evolution of the total energy and its individual contributions, we observe that at the beginning the total energy increases monotonically in time, reflecting the continuous energy injection induced by the hyperbolic-tangent ramp-up of the barrier velocity. Most interestingly, while no remarkable energetic features are observed in the sub-critical regime, for $v_b^{max}\geq v_c$ each vortex nucleation event is associated to a change in the slope of the total energy $E_\mathrm{tot}$ and the classical kinetic energy $E_k^\mathrm{class}=\int_\Omega d\Omega \frac{1}{2}\rho v^2$ [Fig.~\ref{fig:energy} (top) and (bottom), respectively], followed by a sharper increase and a subsequent quasi-plateau.
When a vortex is nucleated and leaves the low-density region created by the barrier traveling toward the bulk, it increases the kinetic energy (and hence the total energy) of the BEC as the density surrounding the vortex ring is larger. This increase continues until the vortex reaches its asymptotic propagation regime in the bulk. Thereon, the total energy exhibits a quasi-plateau until a new vortex nucleation event occurs, producing a further increase.

\section{Destabilization of vortex ring}

\begin{figure}[h]
    \centering
    \includegraphics[width=1.0\columnwidth]{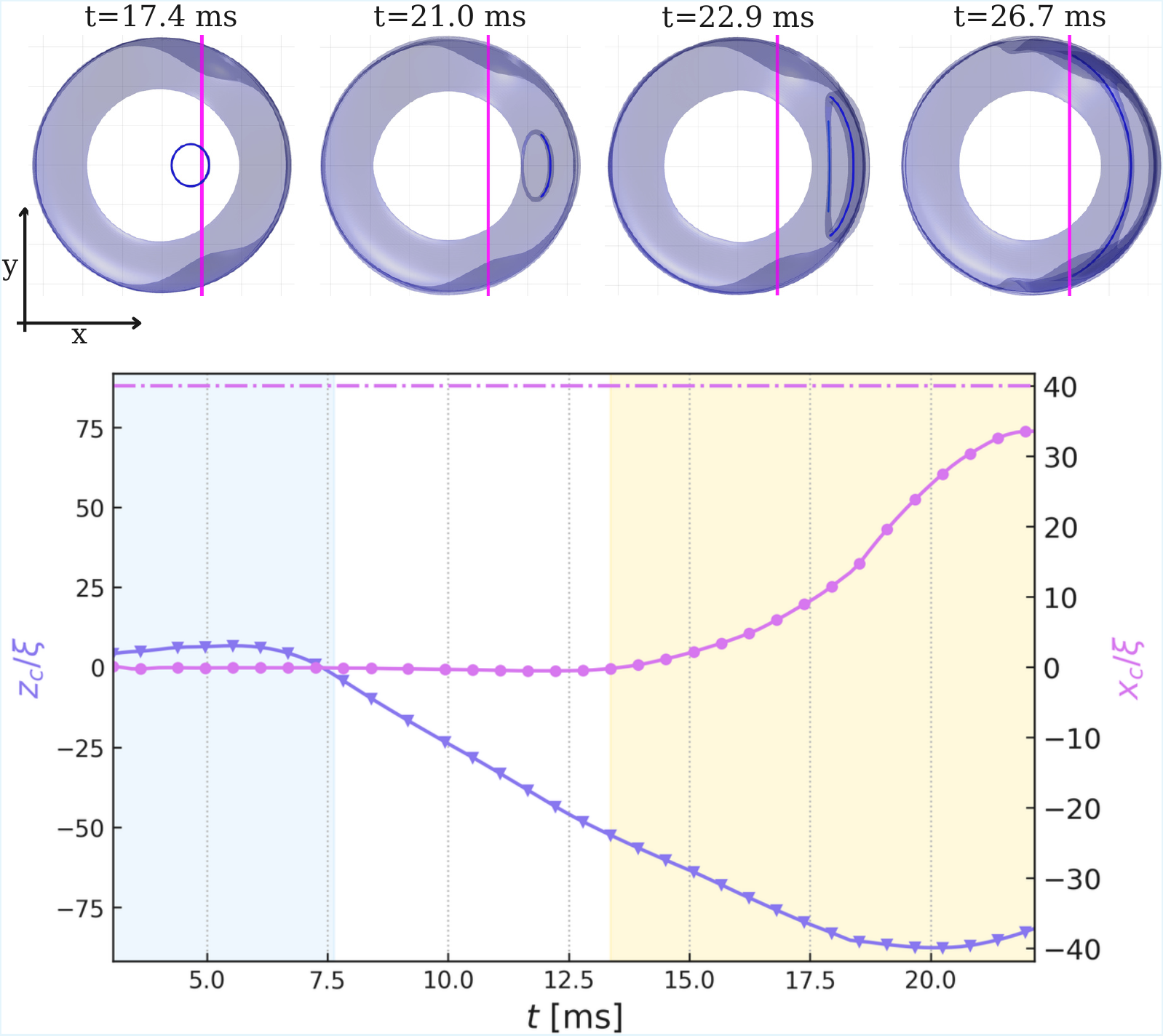}
    \caption{Triggering of vortex ring destabilization. Top panel: BEC density iso-surfaces at four different times viewed frontally along the $z$ axis; the local vertical potential $V(\mathbf{x})$ is represented as a pink column. Bottom panel: Time evolution of the vortex centre's $x$ (pink circles) and $z$ (blue triangles) positions. After leaving the blue area \textemdash where the vortex interacts with the moving barrier\textemdash the vortex exhibits its usual linear motion along the $z$ axis, with no motion in the $x$ direction. Upon entering the pink area \textemdash where density is strongly affected by the vertical obstacle\textemdash the vortex centre's $x$ coordinate starts increasing linearly, showing that the vortex is bending toward the obstacle. Eventually, the $x$ velocity increases while the $z$ velocity decreases, until the vortex approaches the trap boundary. The dash-dot line indicates the trap Thomas--Fermi radius.}
    \label{fig:shifted_obstacle}
\end{figure}

In order to trigger the destabilization of the vortex ring, we employ one of the laser beams used for the excitation of Kelvin waves in Fig.~\ref{fig:fig4}, but position it off centre. The local vertical potential representing the laser beam is thus 
$\displaystyle V(\mathbf{x})=V_c \exp{\{-\left[(x-x_c)^2+(z-z_c)^2\right]/2\sigma_c^2\}}$, where $\sigma_c=10\xi$, $V_c/\mu=3/8$ and $z_c=-10$ as in the previous case, while
$x_c=2$. The vortex ring reaches the region where the density is significantly affected by the local potential at $t \approx 13.5$ ms (yellow area in Fig.~\ref{fig:shifted_obstacle}, top). From that time onward, the ring starts bending toward the beam until crashing onto the boundary. This behaviour is shown in Fig.~\ref{fig:shifted_obstacle} both visually (top) and by noting that after $t \approx 13.5$ ms the vortex acquires a linear motion in the positive $x$ direction, while its motion along the $z$ axis of the cylinder is eventually suppressed, leading to its final impact with the boundary (bottom). We observe that the strong density depletion at the boundary leads to a substantial enlargement of the vortex as it approaches the edge.

\newpage




\clearpage

\appendix


\clearpage

\onecolumngrid
\newpage
\begin{center}
 \textbf{\large Supplementary Material}
\end{center}
\twocolumngrid

\setcounter{figure}{0}
\setcounter{table}{0}
\setcounter{equation}{0}
\setcounter{section}{0}

\renewcommand{\thefigure}{S\arabic{figure}}
\renewcommand{\thetable}{S\arabic{table}}
\renewcommand{\theequation}{S\arabic{equation}}

\newcommand{\rrangle}{\rangle\!\rangle}
\newcommand{\llangle}{\langle\!\langle}
\newcommand{\xit}{\tilde{\xi}}
\newcommand{\rot}{\tilde{r}_0}

\section*{Theoretical and numerical framework}
The system under investigation is a dilute Bose--Einstein condensate in the shape of a cylinder, subject to a harmonic trapping given by an external potential of the form 
\begin{equation}
    \label{eq:simulation_potential}
    V_h(\mathbf{r}) = \frac{1}{2} m (\omega_\perp^2(x^2+y^2) + \omega_z^2z^2) \qquad \omega_z<\omega_\perp
\end{equation}
with trapping frequencies $\omega_\perp=2\pi\cdot 150\ Hz$ and $\omega_z=2\pi\cdot 15\ Hz$ which make the condensate acquire radial symmetry in the $x$-$y$ plane and elongate in the $z$-direction.\\
The dynamics of the system is described by the Gross--Pitaevskii model. We choose the $x$-$y$ frequency $\omega_\perp$ as the reference trapping frequency. The relevant units for the problem thus become
\begin{equation}
    \label{eq:relevant_units}
    l_\perp = \sqrt{\frac{\hbar}{m\omega_\perp}} \qquad \epsilon_\perp = \hbar \omega_\perp \qquad \tau_\perp = \frac{1}{\omega_\perp}
\end{equation}
respectively for length, energy and time. In our case, $l_\perp=2.37\ \mu m$, $\epsilon_\perp=9.934\cdot10^{-32}\ J$ and $\tau_\perp=1.061\ ms$. In these units, we let the BEC wavefunction $\Psi$ evolve according to the time-dependent Gross--Pitaevskii equation. Rescaling the equation by the number of particles $N$, the GPE is expressed in the following form:
\begin{equation}
    \label{eq:GPE_scaled}
    i\frac{\partial\Psi}{\partial t} = -\frac{1}{2}\nabla^2\Psi + V\Psi + C |\Psi|^2\Psi - \mu\Psi
\end{equation}
where $C=4\pi N a_s/l_\perp$ is the rescaled interaction strength, with $a_s$ the s-wave scattering length. The normalization condition for the wavefunction features
\begin{equation}
    \label{eq:normalization_scaled}
    \int{d\mathbf{r}|\Psi|^2} = 1
\end{equation}

\subsection*{Trapping potential effects on ground-state density profile}
\begin{figure}[htp]
    \centering
     \centering
    \includegraphics[width=1.0\columnwidth]{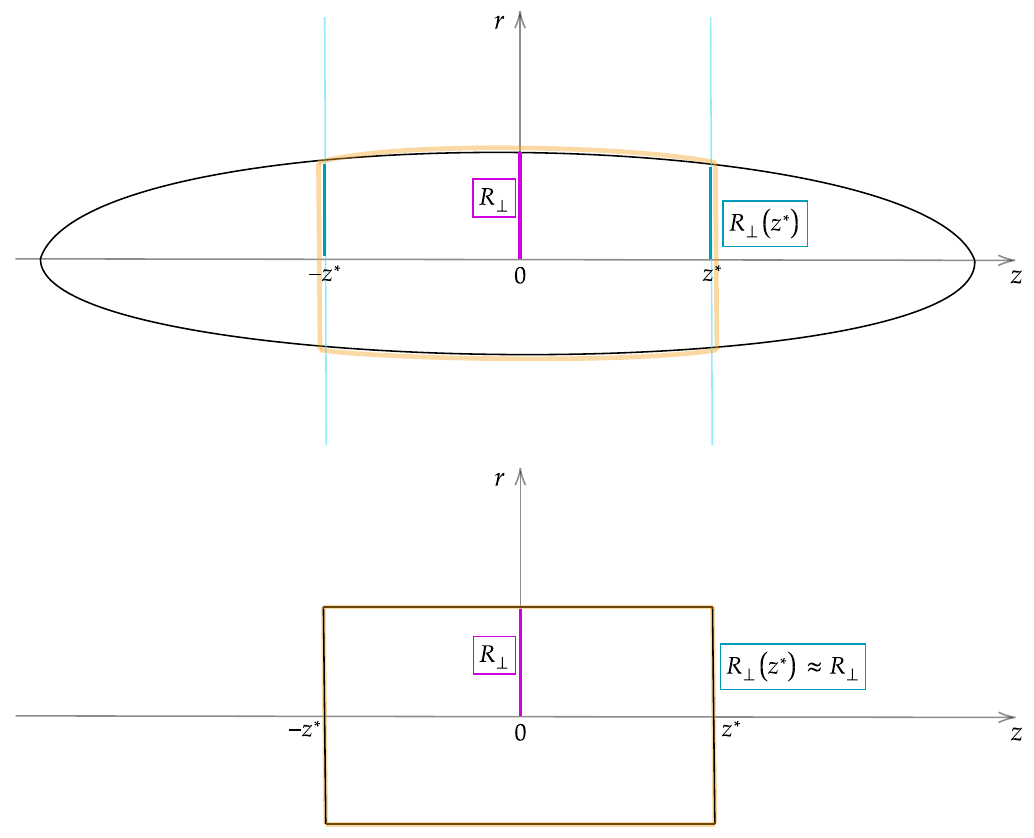}
    \caption{Two-dimensional visualization of the density profile of the condensate under the action of the trapping potential described by Eq.~\eqref{eq:simulation_potential}. Top panel: The black ellipsoid represents the original shape given by the cigar-like trapping potential. The orange-delimited region is the region we focus on. Bottom panel: Delimited in orange, the approximation of the central region of the ellipsoid as a cylinder.} 
    \label{fig:traps}
\end{figure}
The lowest-energy solution of Eq.~\eqref{eq:GPE_scaled} represents the ground state of our cylindrical condensate, which satisfies the time-independent Gross--Pitaevskii equation:
\begin{equation}
    \label{eq:GPE_time_ind}
    \mu\Psi = -\frac{1}{2}\nabla^2\Psi + V\Psi + C |\Psi|^2\Psi
\end{equation}
The ground-state density profile of this configuration can be found by applying the Thomas--Fermi approximation: in the limit of strong interactions, the kinetic term in Eq.~\eqref{eq:GPE_time_ind} is negligible with respect to the interaction term. This leads to the so-called Thomas--Fermi particle density profile, thus described by
\begin{equation}
    \label{eq:density_gs_simulation}
    n = |\Psi|^2 = 
    \begin{cases}
        \frac{\mu-V}{C} \qquad &V<\mu\\
        0 &\mathrm{elsewhere}
    \end{cases}
\end{equation}
\par The region of space where the density in Eq.~\eqref{eq:density_gs_simulation} does not vanish imparts its shape to the condensate itself. The trapping potential in Eq.~\eqref{eq:simulation_potential} defines an ellipsoidal Thomas–Fermi surface. Choosing the symmetry axis along the $z$ direction, the transverse confinement depends on $z$ due to the harmonic trapping along this axis. To obtain an effectively cylindrical geometry, we restrict the system to the central region of the ellipsoid by considering $z \in [-z^*, z^*]$. As shown in Fig.~\ref{fig:traps}, the resulting condensate has the shape of a cylinder of length $2z^*$ and radius $R_\perp = \sqrt{2\mu}$, determined by the Thomas–Fermi profile. Therefore, the trapping potential becomes
\begin{equation}
    \label{eq:cyl_trap}
    \begin{split}
    V_h(x,y,z) = \frac{1}{2}r^2 \qquad &r=\sqrt{x^2+y^2}\ , \\ &z\in[-z^*,z^*] 
    \end{split}
\end{equation}

\subsection*{System parameters}
The particle density discussed in the previous paragraph (Eq.~\eqref{eq:density_gs_simulation}) comes from a wavefunction normalized to unity. This leads to the following constraint:
\begin{equation}
    \label{eq:mu_constraint}
        1 = \int_{\mathcal{C}}d\mathbf{r}\ n(\mathbf{r}) =\frac{\mu}{C}\int_{\mathcal{C}}d\mathbf{r}\ \biggl(1-\frac{V(\mathbf{r})}{\mu}\biggr)
\end{equation}
where $\mathcal{C}$ is the cylindrical Thomas--Fermi profile of the condensate. The resolution of the integrals in Eq.~\eqref{eq:mu_constraint} yields a relation that links the number of particles $N$, the characteristics of the considered atomic species (i.e. $m$ and $a_s$) and the trapping frequencies $\omega_z$ and $\omega_\perp$ to the chemical potential:
\begin{equation}
    \label{eq:mu_cyl}
    \mu = \biggl(\frac{\sqrt{2}a_sR_\perp N}{z^* l_\perp}\biggr)^{2/5}
\end{equation}
This relation allows us to compute the chemical potential of the cylindrical condensate knowing the specific characteristics of the system and the trap.
\par Tab. \ref{tab:parameters_table} reports the values of the condensate parameters used for all the simulations in this article.
\begin{table}[h]
    \centering
    \begin{tabular}{|c|c|}
        \hline
        \textbf{Parameter} & \textbf{Value} \\
        \hline
        $\mu$ & $1.987\cdot 10^{-30}\ J$ \\
        \hline
        $\omega_z$ & $2\pi\cdot 15\ Hz$ \\
        \hline
        $\omega_\perp$ & $2\pi\cdot 150\ Hz$ \\
        \hline
        $R_\perp$ & $14.99\ \mu m$ \\
        \hline
        $z^*$ & $37.48\ \mu m$ \\
        \hline
        $N$ & $1.42\cdot10^5$ \\
        \hline
        $C=g/N$ & $1.051\cdot 10^{-42}\ J\cdot m^3$ \\
        \hline
    \end{tabular}
    \caption{Physical parameters values used in the simulations. These values are based on equilibrating a realistic trapped BEC of $^6$Li atoms and are chosen to avoid a destructive increase in computational cost. This system is characterized by a bulk sound speed $c_\infty=7.065\ mm/s$. The reported length of the cylindrical trap $z^*$ is the minimum length used in the simulations; the values of $N$ and $C$ are tailored for such cylinder.}
    \label{tab:parameters_table}
\end{table}

\section*{Introduction of the moving barrier}
To achieve vortex nucleation, a repulsive obstacle is introduced in the previously described cylindrical condensate. Specifically, this obstacle consists of a Gaussian barrier transverse to the axis of the cylinder. It is represented by a time-dependent potential in the form
\begin{equation}
    \label{eq:barrier}
    V_b(x,y,z,t) = V_0e^{-(z-z_b(t))^2/(2\sigma_b^2)}
\end{equation}
where $V_0$, $\sigma_b$ and $z_b(t)$ are, respectively, the height of the barrier, its width and the position of its center at time $t$. In an experimental context, the counterpart of this potential barrier would be a blue-detuned laser beam characterized by a width $\sigma_b$, typically ranging from $5\xi$ to $10\xi$.
\par In order to reduce the generation of sound caused by the motion of the barrier, the center of the barrier moves 
with velocity $v_b(t)$ given by the following relation
\begin{equation}
    \label{eq:vel_tanh}
    v_b(t) = v_b^\mathrm{max}\mathrm{tanh}\biggl(\frac{t}{\tau_b}\biggr)
\end{equation}
where $\tau_b$ identifies the characteristic time scale for the onset of the barrier’s motion, set to $\tau_b=5.305\ ms$ in these simulations. The value of $\tau_b$ regulates the smoothness with which the barrier's velocity achieves the asymptotic value $v_b^\mathrm{max}$: varying $\tau_b$ slightly changes the values of $v_c$ but does not modify the picture described.  

\subsection*{Effect of the barrier on ground-state density profile}
\begin{figure}[htp]
    \centering
    \includegraphics[width=0.8\columnwidth]{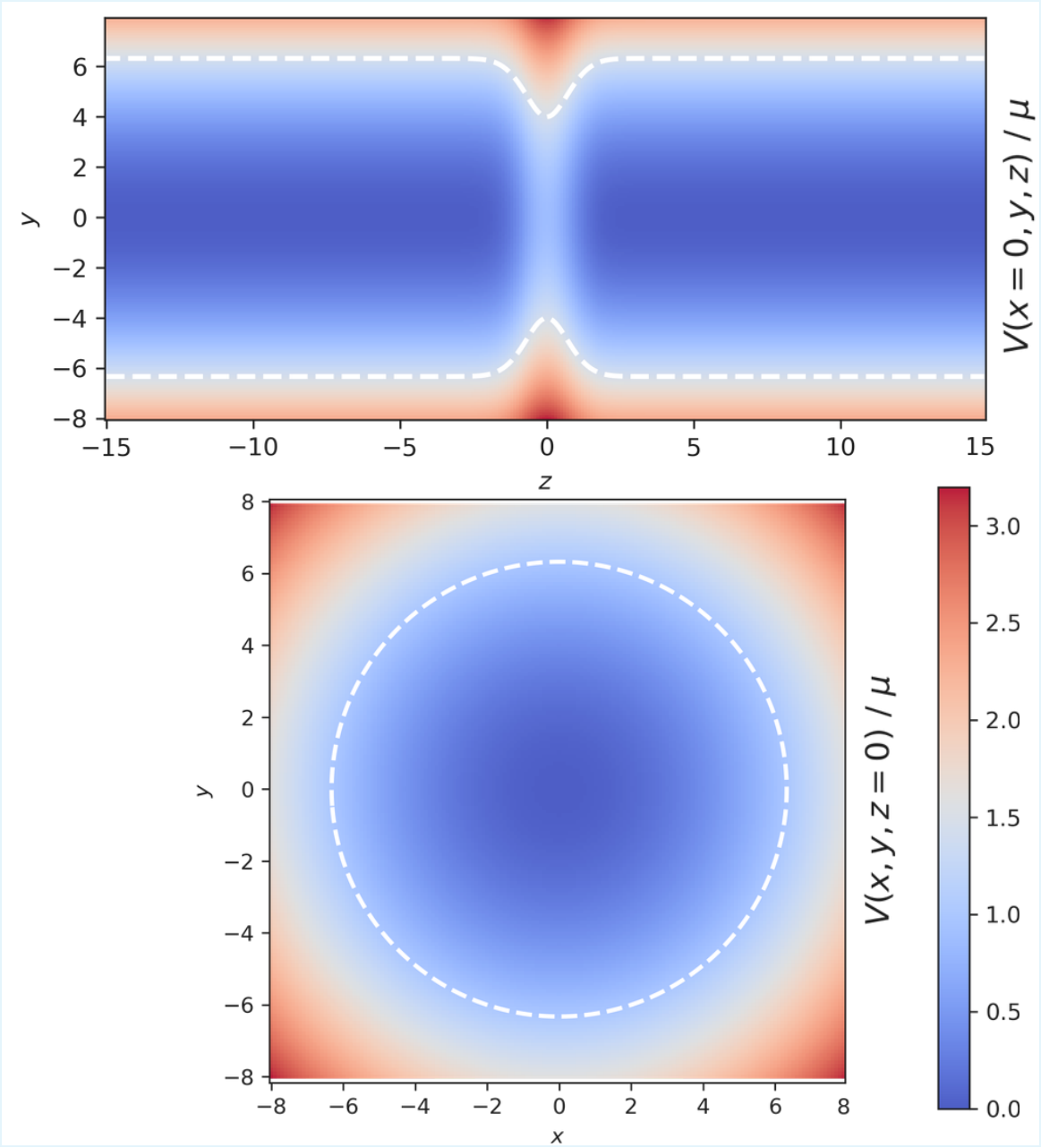}
    \caption{Two-dimensional visualization of the external potential $V/\mu$ projected on plane $x=0$ (top) and $z=0$ (bottom). Here we consider the barrier centered in $z_0=0$. The white dashed line marks the points where $V/\mu=1$, i.e. the projections of the Thomas--Fermi surface.} 
    \label{fig:trap+barr}
\end{figure}

The initial form of the barrier corresponds to evaluating Eq.~\eqref{eq:barrier} at $t=0$. The external potential consists of two contributions, the cylindrical harmonic trap (Eq.~\eqref{eq:cyl_trap}) and the Gaussian barrier (Eq.~\eqref{eq:barrier}):
\begin{equation}
    \label{eq:extpotbarrier}
    \begin{split}
    V(x,y,z) &= V_h(x,y,z) + V_b(x,y,z) =\\
    &=\frac{1}{2}(x^2+y^2) + V_0e^{-(z-z_0)^2/(2\sigma_b^2)}
    \end{split}
\end{equation}
with $z\in[-z^*,z^*]$. The total external potential $V$ at $t=0$ is shown in Fig.~\ref{fig:trap+barr}. The Thomas--Fermi approximation yields the same analytical form for the density profile as Eq.~\eqref{eq:density_gs_simulation}, thus the Gaussian barrier acts as a radially symmetric transverse constriction applied on the cylinder, squeezing the condensate and reducing its transverse radius at each $z$. The condensate boundary is now a cylinder with a $z$-dependent transverse radius, obtained by imposing the limiting condition $V=\mu$. The cylinder is maximally constricted at the center of the barrier (i.e. $z = z_0$), where the barrier reaches its maximum height. This maximum height value determines the minimum transverse radius:
\begin{equation}
    \label{eq:R_perp_barrier_minimum}
    R_{\perp_b} = R_\perp(z_0) = \sqrt{2(\mu - V_0)}
\end{equation}

\subsection*{Gross--Pitaevskii equation in the moving reference frame}
If the barrier moves with velocity $\mathrm{w}$ along the cylinder axis, it may be convenient to switch to the co-moving coordinate system where the barrier is stationary.  
In this frame the Gross–Pitaevskii equation acquires an additional term compared to Eq.~\eqref{eq:GPE_scaled}, becoming
\begin{equation}
    \label{eq:GPE_moving_frame}
    i\frac{\partial\Psi}{\partial t} = -\frac{1}{2}\nabla^2\Psi + i\mathrm{w}\frac{\partial\Psi}{\partial z} + V\Psi + C |\Psi|^2\Psi - \mu\Psi
\end{equation}

The external potential expressed in the new coordinates reads as follows:
\begin{equation}
    \label{eq:extpot_movingframe}
    V(x,y,z) = \frac{1}{2}(x^2+y^2) + V_0e^{-z^2/(2\sigma_b^2)}
\end{equation}

\section*{Exploiting the cylindrical symmetry}
The system under investigation exhibits cylindrical symmetry, with no dependence on the azimuthal angle $\theta$. Consequently, the wavefunction $\Psi(x,y,z)$ may be written as $\Psi(r,\theta,z)=\Psi(r,z)$. In this new coordinate system, the Gross--Pitaevskii equation becomes
\begin{equation}
    \begin{split}
        i\frac{\partial\Psi(z,r)}{\partial t} = &-\frac{1}{2}\biggl(\frac{\partial^2}{\partial z^2} + \frac{\partial^2}{\partial r^2} \biggr) \Psi(z,r){-\frac{1}{2r}\frac{\partial\Psi(r,z)}{\partial r}}\\
        & + \biggl(\frac{1}{2}r^2 + V_0 \ e^{-\frac{(z-z_0)^2}{2\sigma_b^2}}\biggr)\Psi(z,r)\\ 
        &+ C|\Psi(z,r)|^2\Psi(z,r) -\mu\Psi(z,r),
    \end{split}
\end{equation}
where the Laplacian has been expressed in cylindrical coordinates, with the azimuthal derivative vanishing due to symmetry, and the trap and moving barrier potentials are explicitly written in these coordinates. The term proportional to $1/r$ requires special attention, since it diverges at $r=0$. This singularity can be removed by imposing regularity of the wavefunction on the symmetry axis, namely $\partial_r\Psi(z,0)=0$. As a result, the three-dimensional problem has been effectively reduced to a two-dimensional problem, leading to a significant advantage in computational time. 
\par This dimensionality reduction can be exploited only if cylindrical symmetry is maintained throughout the full dynamics. We verified that the nucleation dynamics, including the determination of the critical barrier velocity, is the same as in the cartesian coordinate system. We therefore carried out the simulations of the symmetrical system in cylindrical coordinates, and subsequently reconstructed vortex position using an algorithm based on two-dimensional pseudo-vorticity \cite{Stagg-code2d}.

\section*{Triggering of Kelvin waves and vortex ring destabilization}
In order to generate excitations on the vortex rings previously nucleated by the moving barrier, it is necessary to introduce obstacles that break the symmetry of the system. To this end, we employ two laser beams parallel to the $y$ direction, modeled as two local vertical potentials centered at $x_i=\pm 3$, $z_c=-10$ with the following form:
\begin{equation}
    \label{eq:vertical_potentials}
    V_i(\mathbf{r})=V_c e^{-\frac{(x-x_i)^2+(z-z_c)^2}{2\sigma_c^2}}
\end{equation}
with intensity $V_c=3/8\mu$ and width $\sigma_c=10\xi$. When the vortex ring passes through the region where the vertical obstacles significantly deplete the density, it stretches along the $x$ direction and becomes elliptical. This deformation triggers a Kelvin wave of mode $m=2$ that can be observed in the oscillatory behaviour of the semiaxes $a$ and $b$ of the $x$-$y$ planar projection of the vortex ring. A fit of this oscillatory behaviour confirms the nature of the excitation (Fig.4 of main manuscript, top, right inset), showing good agreement with the theoretical frequency for Kelvin waves in the limit where the ring radius is larger than the core size, given by
\begin{equation}
    \label{eq:frequencyKW}
    \omega(k)\approx \frac{\kappa k^2}{4\pi}\biggl[\mathrm{ln}\biggl(\frac{2}{k\xi}\biggr)-\gamma\biggr]
\end{equation}
where $\gamma=0.5772$ and $k=2/R$ for mode $m=2$, $R$  being the radius of the vortex ring before the symmetry breaking occurs \cite{Pocklington-1895}.
\par
\begin{figure}[htp]
    \centering
    \includegraphics[width=0.8\columnwidth]{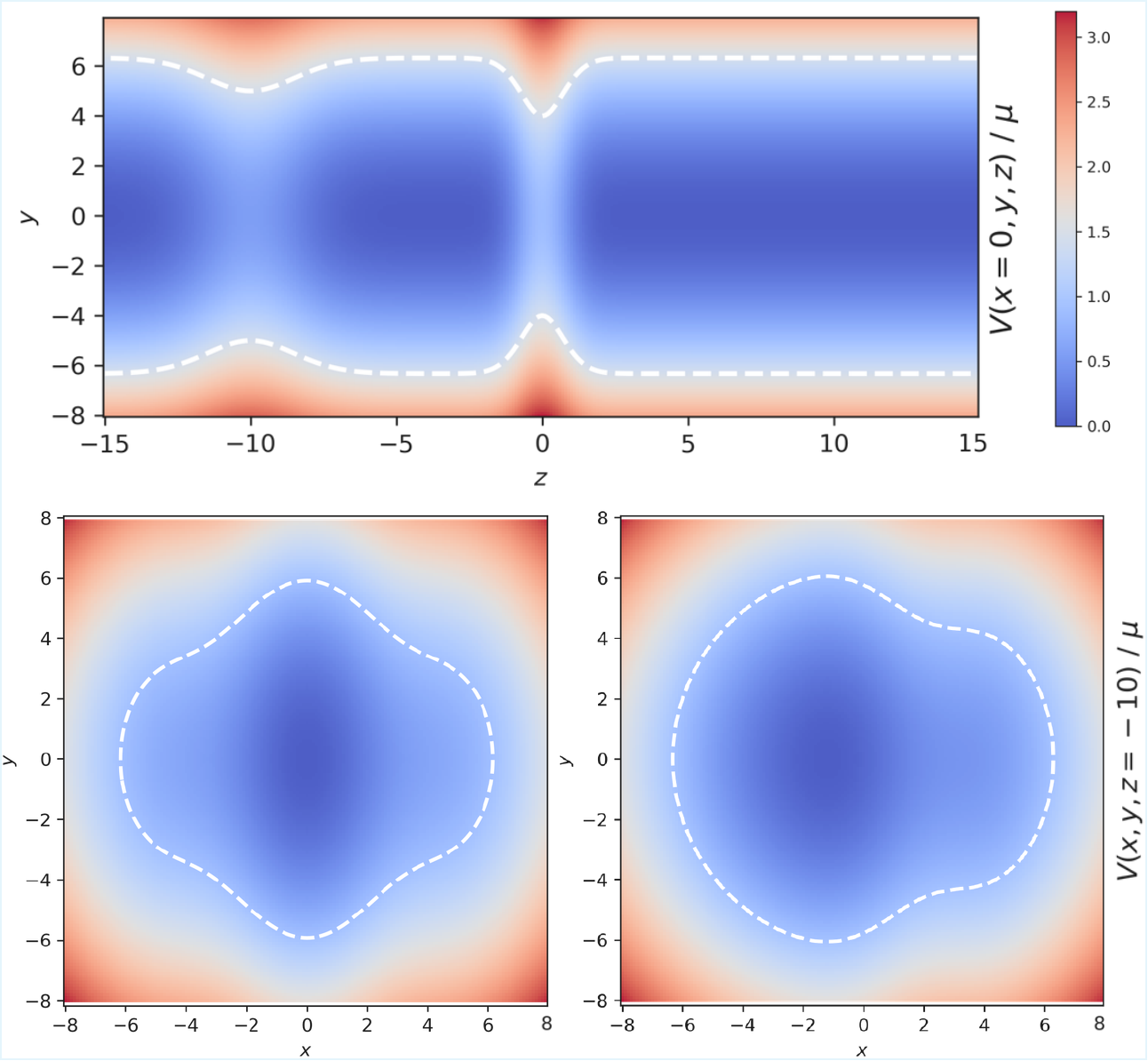}
    \caption{Two-dimensional visualization of the external potential $V/\mu$ including the vertical obstacles, projected on plane $x=0$ (top) and $z=-10$ (bottom). The $z=-10$ plane is displayed for both the case with two vertical potentials (left) and a single off-centered potential (right). Since all vertical potentials are located at $z_c=-10$, the $x=0$ projection is nearly identical in both configurations and is shown only once. Here we consider the barrier centered in $z_0=0$. The white dashed line marks the points where $V/\mu=1$, i.e. the projections of the Thomas--Fermi surface.} 
    \label{fig:trap+barr+col}
\end{figure}

 If only one of the two vertical potentials is applied and positioned off-center (Eq.~\eqref{eq:vertical_potentials} with $x_i=2$), it causes the vortex ring to bend toward the potential. Thus, the ring begins to move in the positive $x$ direction, approaching the boundary and expanding in size until it eventually crashes onto it. The total external potential $V$ in the two configurations described above is shown in Fig.~\ref{fig:trap+barr+col}.

\section*{Computational resolutions and periodic boundary conditions}
The definition of the computational grid representing space requires choosing an appropriate spatial resolution. This can be determined by the relevant healing length of the system, which in this case is the non-dimensional healing length on the cylinder axis:
\begin{equation}
    \label{eq:healinglength_cyl}
    \tilde{\xi}_0 =\frac{\xi_0}{l_\perp} =  \frac{1}{l_\perp}\sqrt{\frac{\hbar^2}{2mgn_0}}= \frac{1}{\sqrt{2\tilde{\mu}}}
\end{equation}
For our system, the healing length on the cylinder axis is thus $\xi_0=0.374\ \mu m$. An accurate spatial resolution is ensured by a grid step of $\Delta x\approx\xi_0/3 = 0.14\ \mu m$. 
\par In cartesian coordinates, we impose periodic boundary conditions in all the three spatial directions. The simulation domain is therefore a box $[-L_x, L_x]\times[-L_y, L_y]\times[-L_z, L_z]$, which defines the unit cell repeated periodically. The length in the $z$-direction is $L_z=z^*$, which corresponds to the truncation along the $z$-axis for the cylindrical approximation. As for $x$ and $y$ directions, $L_x$ and $L_y$ must be larger than the transverse radius $R_\perp$ to accommodate the region where the density decays to zero\textemdash density does not suddenly vanish to zero, even though for $r>R_\perp$ it assumes very low values. In this sense, a safe margin is given by a distance of $20\xi_0$. Hence, we set $L_x=L_y=10l_\perp$ and $L_z=15l_\perp$ as the minimum cylinder length. The number of grid points in each direction is then defined as follows:
\begin{equation}
    \label{eq:num_points}
    N_i =\frac{2L_i}{\Delta x}
\end{equation}
which yields $N_x=N_y=340$ and $N_z=512$.
\par In cylindrical coordinates, the simulation domain is $[-L_r,L_r]\times[-L_z,L_z]$ with $L_r=L_x=L_y$; periodic boundary conditions are applied along the $z$ axis, while mirror (Neumann) boundary conditions are imposed in the radial direction.
\par The time resolution is chosen based on the spatial resolution and the structure of the Gross--Pitaevskii equation (see Eq.~\eqref{eq:GPE_scaled}). Since the time-derivative term is comparable to the Laplacian term, a stable numerical integration requires a time step that satisfies $\Delta\tilde{t}<\Delta\tilde{x}$. A value of $\Delta t = \Delta x/4 = 9.549\cdot10^{-4}\ ms$ is sufficient.

\section*{Calculation of Kinetic Energy and Momentum of BEC in presence of a vortex ring}
\begin{figure}[htbp]
    \centering
    \includegraphics[width=0.7\columnwidth]{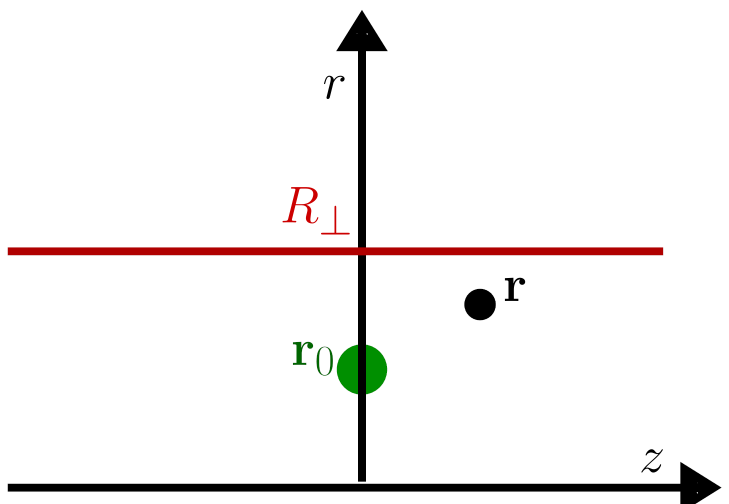}
    \caption{$(z,r)$ cut of the cylindrical condensate. A negative vortex is placed in $\mathbf{r}_0(0,r_0)$ and $\mathbf{r}(z,r)$ 
    is a generic point on the $(z,r)$ plane. The system
    is axisymmetric, hence no dependence on the azimuthal angle $\theta$ is present.}
    \label{fig:channel_sketch}
\end{figure}

We analytically derive expressions for the kinetic energy and momentum of an inhomogeneous BEC confined in a cylindrical harmonic trap, in the presence of a vortex ring with radius $r_0$ and in the absence of any additional barrier (see Fig.~\ref{fig:channel_sketch}).

\subsection*{Energy}
The density in a generic point $\mathbf{r}(z,r)$ is given \cite{jackson-etal-1999} by the following expression
\begin{equation}
    \label{eq:density_approx}
\displaystyle \rho =\rho(z,r) = \rho_0 \left [ \frac{|\mathbf{r}-\mathbf{r_0}|^2}{|\mathbf{r}-\mathbf{r_0}|^2 + \xi^2}\right ] 
\left [ 1-\left ( \frac{r}{R_\perp} \right )^2 \right ] \, \, ,
\end{equation}

\noindent 
where $\rho_0$ is the density on the cylinder axis ($r=0$) and $\xi$ the corresponding healing length. 
The superfluid velocity $\mathbf{v}$ in the local approximation where the velocity is given by 
\begin{equation}
    \label{eq:velocity_approx}
\mathbf{v}=\bm{\kappa} \times (\mathbf{r}-\mathbf{r}_0) /(2\pi|\mathbf{r}-\mathbf{r}_0|^2) \, \, ,
\end{equation}  
where $\bm{\kappa}$ is the oriented vortex line element parallel to the vorticity and magnitude equal to the quantum of circulation
$\kappa=h/m$. The magnitude of the velocity is hence given by
\begin{equation}
    \label{eq:velocity_approx_modulus}
\displaystyle 
v=v(z,r)=\frac{\kappa}{2\pi\sqrt{(r-r_0)^2+z^2}} \, \, .
\end{equation}  

We neglect the quantum pressure term and hence the kinetic energy $E$ of the condensate is hence as follows

\begin{eqnarray}
\label{eq:E_kin}
\displaystyle 
E & = & \int_\mathcal{C}\frac{1}{2}\rho v^2 d\mathbf{r} = \pi \int_0^{R_\perp} \!\!\!\!\! dr \int_{-\infty}^\infty \!\!\!\!\! dz \,\, r \rho(z,r) v^2 (z,r) \nonumber \\[2mm]
  & = & \frac{\rho_0 \kappa^2}{4\pi} \int_0^{R_\perp} \!\!\!\!\! dr \,\,r \left [ 1-\left ( \frac{r}{R_\perp} \right )^2 \right ] \int_{-\infty}^\infty \!\!\!\!\! dz \,\frac{1}{(r-r_0)^2 + z^2 +\xi^2} \nonumber \\[2mm]
  & = & \frac{\rho_0 \kappa^2 R_\perp}{4}\left [ \frac{1}{R_\perp}\int_0^{R_\perp} \!\!\!\!\! dr \,\, \frac{r}{\sqrt{(r-r_0)^2 + \xi^2}} \right. \nonumber \\
  && \left. - \frac{1}{R_\perp^3}\int_0^{R_\perp} \!\!\!\!\! dr \,\, \frac{r^3}{\sqrt{(r-r_0)^2 + \xi^2}} \right ] \nonumber \\
  & = & \frac{\rho_0 \kappa^2 R_\perp}{4} \left \{ \sqrt{(1-\tilde{r}_0)^2 + \tilde{\xi}^2}-\sqrt{\tilde{r}_0^2+\tilde{\xi}^2} \right. \nonumber \\
  && + \rot \left \{ \log \left [ \frac{1-\rot + \sqrt{(1-\rot)^2 + \xit^2}}{-\rot + \sqrt{\rot^2+\xit^2}} \right ] \right \} \nonumber \\
  && -(1-\rot)^2\sqrt{(1-\rot)^2+\xit^2} \nonumber \\
  && + \rot^2\sqrt{\rot^2+\xit^2}+\frac{2}{3}\left [ (1-\rot)^2+\xit^2\right ]^{3/2} -\frac{2}{3}\left ( \rot^2+\xit^2 \right )^{3/2} \nonumber \\
  && -3\rot\left [ \frac{1-\rot}{2}\sqrt{(1-\rot)^2+\xit^2} +\frac{\rot}{2}\sqrt{\rot^2+\xit^2}\right ] \nonumber \\
  &&  -3\rot^2 \left [ \sqrt{(1-\rot)^2+\xit^2} -  \sqrt{\rot^2+\xit^2}   \right ] \nonumber \\
  && \left. - \rot^3 \left \{ \log \left [ \frac{1-\rot + \sqrt{(1-\rot)^2 + \xit^2}}{-\rot + \sqrt{\rot^2+\xit^2}} \right ] \right \} \right \} \, \, ,
\end{eqnarray}

\noindent 
where $\mathcal{C}$ is the cylinder region occupied by the BEC in the Thomas--Fermi limit, $\rot=r_0/R_\perp$ and $\xit=\xi/R_\perp$.
The dependence of $E$ as a function of the radius of the ring is reported in the main manuscript, Fig.4 (bottom, inset).

\subsection*{Momentum}

The axial component of the velocity $v_z$ induced at $\mathbf{r}(z,r)$ by the negative vortex placed in $\mathbf{r}_0(0,r_0)$ is given 
by the following expression 
\begin{equation}
\label{eq:v_z}
\displaystyle v_z=\frac{\kappa (r-r_0)}{2\pi [ (r-r_0)^2 + z^2 ]}
\end{equation}

Thus, starting from Eq. (4) in the End Matter of the main manuscript, we obtain the following results
\begin{eqnarray}
\label{eq:p_z}
\displaystyle 
p_z & = & \frac{1}{2 i}\int_\mathcal{C} \left ( \Psi^*\frac{\partial \Psi}{\partial z}-\Psi\frac{\partial \Psi^*}{\partial z}\right ) d\mathbf{r} = 
    \int_\mathcal{C} \rho(z,r) v_z(z,r) d\mathbf{r} \nonumber \\
    & = & \kappa \rho_0 \int_0^{R_\perp} \!\!\!\!\! dr \,\,r (r-r_0)\left [ 1-\left ( \frac{r}{R_\perp} \right )^2 \right ] \int_{-\infty}^\infty \!\!\!\!\! dz \,\frac{1}{(r-r_0)^2 + z^2 +\xi^2} \nonumber \\[2mm]
    & = & \kappa \rho_0 \pi R_\perp^2 \int_0^1 \!\!\!\!\! dr \,\,r (1-r^2) \frac{r-r_0}{\sqrt{(r-r_0)^2 + \xi^2}} \nonumber \\
    & = & \kappa \rho_0 \pi R_\perp^2 \left \{ \sqrt{(1-\rot)^2+\xit^2}\left [-\frac{1}{2}(1+\rot) +\frac{7}{2}\rot^2-\frac{5}{2}\rot^3\right ] \right. \nonumber \\
    && +\sqrt{\rot^2+\xit^2}\left ( -\frac{1}{2}\rot+\frac{3}{2}\rot^3\right ) + \left [ (1-\rot)^2+\xit^2 \right ]^{3/2} \left ( \frac{3}{4}+\frac{5}{4}\rot \right ) \nonumber \\
    && \left. -\frac{5}{4}\rot\left ( \rot^2+\xit^2\right )^{3/2} \right \} \,\, .
\end{eqnarray}

The dependence of $p_z$ as a function of the radius of the ring is reported in the End Matter of the main manuscript in Fig.5.

\section*{Calculation of Kinetic Energy and Momentum in a classical homogeneous 2D channel}
In classical two-dimensional inviscid fluid dynamics, in the circumstance where the flow is incompressible and irrotational, 
a very useful framework can be used to describe the features of fluid's motion, namely complex potentials \cite{newton-2001}.
According to this formulation, we can define a complex function $\Omega(z)$ in the complex plane $z=x+\rm i y$ as follows:
\begin{equation}
\label{eq:Omega_general}
\displaystyle \Omega(z):= \varphi(x,y) + \rm i \psi(x,y) \, \, ,
\end{equation}
where $\varphi(x,y)$ is the velocity potential such that the velocity $\mathbf{v}(x,y)=\nabla \varphi$ and $\psi(x,y)$ is the streamfunction such 
that the velocity can also be expressed as $\mathbf{v}(x,y)=\nabla \times \bm{\psi}$, where $\bm{\psi}= (0,0,\psi(x,y)$. This construction of $\Omega$
implies that the complex potential is an analytical function of the complex variable $z$, its derivative being the complex velocity $v(z)$, \textit{i.e.}
\begin{equation}
\label{eq:complex_potential_general}
\displaystyle v(z) := \frac{d\Omega}{dz} = v_x - i v_y \,\, .
\end{equation}
In this formulation, the complex potential of a vortex placed in $z_k$, having circulation $\kappa$ in an unbounded fluid, is as follows
\begin{equation}
\label{eq:Omega_vortex}
\displaystyle \Omega(z)= -\text{sgn} (z_k)\frac{\rm i \kappa}{2\pi} \log ( z - z_k)\, \, ,
\end{equation}
where $\text{sgn}(z_k)$ is the sign of the vortex, $\pm 1$ for positive (anti-clockwise) / negative (clockwise) rotational induced flows.

The complex potential-based formulation is very useful as from the knowledge of the complex potential $\Omega_\mathcal{H}(z)$ of a vortex placed in a 
simply connected region
(\textit{e.g.} the imaginary upper half-plane $\mathcal{H}=\{ z\in\mathbb{C} : \Im\text{m} \; z > 0 \}$) via a conformal map $z=f(\zeta)$ it is possible to determine directly the 
complex potential $\Omega_\mathcal{C}(\zeta)$ of a vortex placed in another simply connected region, as for instance an infinite channel  
$\mathcal{C}=\{ \zeta\;\in\;\mathbb{C}\; :\;  0 < \Im\rm m \; \zeta < 2D \} \subsetneq \mathbb{C}$, $D$ being the channel half-width.  
Combining $\Omega_\mathcal{H}(z)$ and $z=f(\zeta)$, we obtain \cite{Galantucci_Sciacca_Parker_Baggaley_Barenghi_2021}
\begin{equation}
\label{eq:Omega_vortex_channel}
\displaystyle \Omega_\mathcal{C}(\zeta)=\Omega_\mathcal{H}(f(\zeta))=-\text{sgn}(\zeta_k)\frac{i\kappa}{2\pi}
\log{\left(\frac{1-e^{^{-\frac{\pi}{2D}(\zeta-\zeta_k)}}}{1-e^{^{-\frac{\pi}{2D}(\zeta-\zeta_k^*)}}}\right)} \, \, ,
\end{equation}
for a vortex is placed at $\zeta_k$ in the channel $\mathbb{C}$ ($^*$ indicates complex conjugation).

\begin{figure}[h]
    \centering
    \includegraphics[width=1.0\columnwidth]{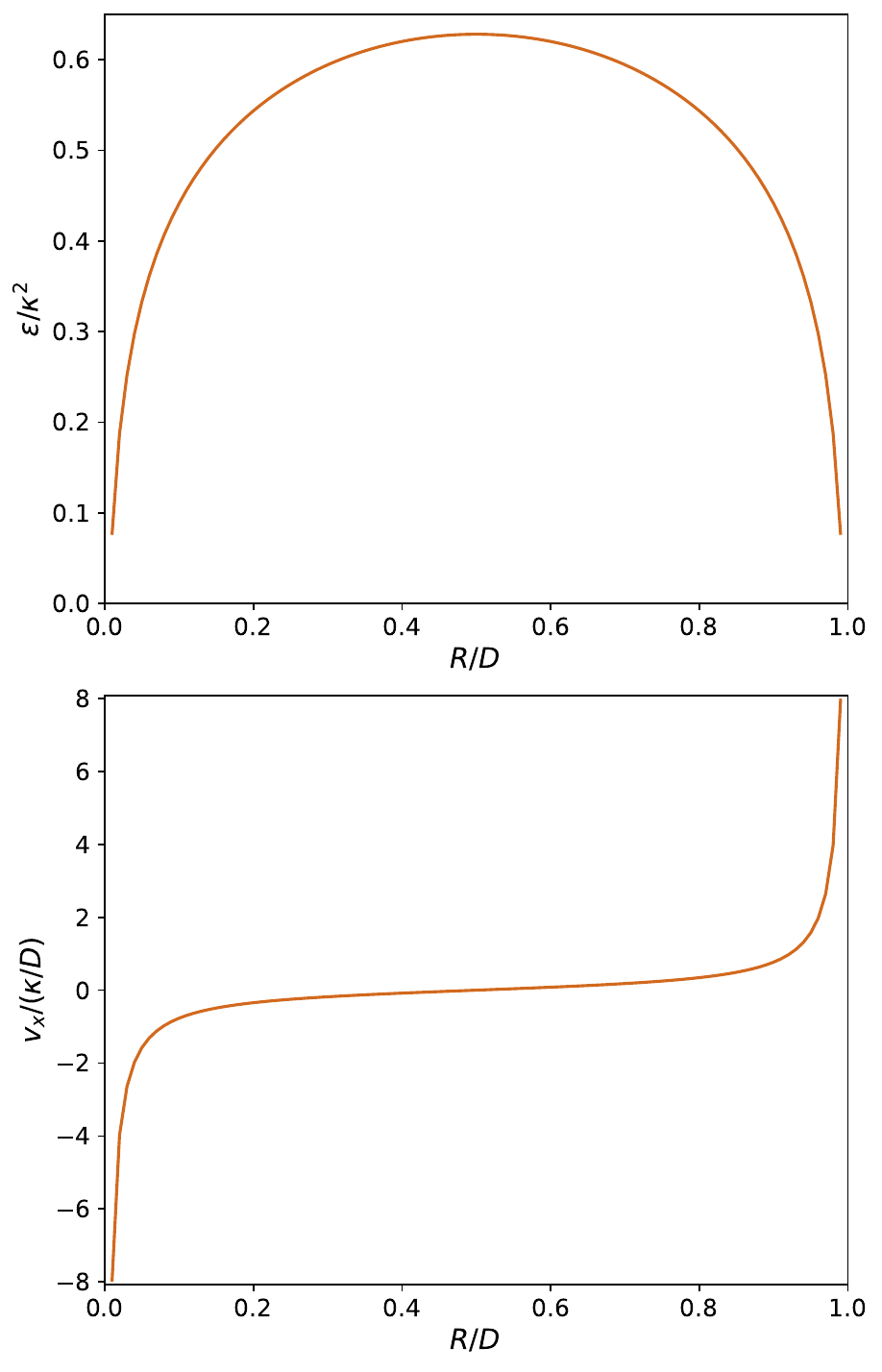}
    \caption{Top panel: Energy per unit length $\varepsilon$ as a function of the radius $R$ of the vortex--antivortex pair in a two-dimensional homogeneous channel. Bottom panel: Velocity $v_x$ as a function of the radius $R$ of the vortex--antivortex pair in a two-dimensional homogeneous channel.}
    \label{fig:channel}
\end{figure}

Without loss of generality, we set the density $\rho=1$ and the channel half-width $D=1$. A vortex--antivortex pair (the two-dimensional equivalent of a vortex ring in three 
dimensions) is placed at $\zeta_\pm=\rm i (1 \mp R)$, where $\zeta_\pm$ correspond to positive/negative vortex and $0<R<1$ is the `radius' of the pair.

The energy per unit length $\varepsilon$ is given by the following relation \cite{Galantucci_Sciacca_Parker_Baggaley_Barenghi_2021}
\begin{eqnarray}
\label{eq:energy_channel}
\displaystyle  \varepsilon & = & \int \frac{1}{2} v^2 d\mathbf{r} = \frac{\kappa}{2}\sum_{k=1}^2 \text{sgn}(\zeta_k) \psi(\zeta_k) \nonumber \\
                           & = & \frac{\kappa^2}{4\pi}\log \left [\frac{1}{\pi^2} | (1-e^{-i\pi R})(1-e^{-i\pi (1+R)}) \times \right. \nonumber \\
                               && \left. (1-e^{i\pi R})(1-e^{-i\pi (1-R)}) | \right ] \, \, ,
\end{eqnarray}

while the axial momentum per unit length $p_x$ is given by ($x$ being the coordinate parallel to the channel walls)
\begin{equation}
\label{eq:momentum_channel}
\displaystyle p_x=\int v_x d\mathbf{r}= \kappa \sum_{k=1}^2 \text{sgn}(\zeta_k) \Im\text{m}(\zeta_k) = -2\kappa R \,\, .
\end{equation}

\noindent
Both quantities $\varepsilon$ and $p_x$ are constant with respect to time, as no energy losses are included in the system (the fluid is inviscid) and the
system is geometrically invariant with respect to translations in the $x$ direction. The dependence of the energy per unit length $\varepsilon$ on the radius $R$
is reported in Fig.~\ref{fig:channel} (top) where we observe that the curve is symmetrical with respect to $R=0.5$. This symmetry, together with the fact that 
$\partial p_x/\partial R = -2\kappa$ is constant with respect to $R$, implies that 
$\displaystyle v_x=\frac{\partial E}{\partial p_x}=\frac{\partial E/\partial R}{\partial p_x/\partial R} = -\frac{1}{2\kappa}  \frac{\partial E}{\partial R}$ is
antisymmetrical with respect to  $R=0.5$ where it vanishes and reverses its sign. This behaviour of the velocity $v_x$ of the vortex--antivortex pair 
can also be directly calculated from Eq.~(\ref{eq:Omega_vortex_channel}). In fact, by deriving $\Omega_\mathcal{C}$ with respect to $z$ and 
employing the superposition principle, we obtain the following expression for $v_x$ 

\begin{equation}
\label{eq:velocity_channel}
\displaystyle v_x=\frac{\kappa}{8D}\left [ \frac{1}{\tan \left ( \frac{\pi R}{2}\right )} - \tan \left ( \frac{\pi R}{2}\right )\right ]
\end{equation}

whose behaviour with respect to $R$ is reported in Fig.~\ref{fig:channel} (bottom), showing the described
antisymmetrical behaviour.



\end{document}